\documentclass[aps,twocolumn,pra,superscriptaddress,amsmath,amssymb]{revtex4-1}
\usepackage{dcolumn}
\usepackage{bm}
\usepackage{amsmath}
\usepackage{txfonts}
\usepackage[T1]{fontenc}
\usepackage{xspace}
\usepackage{ulem}
\usepackage{braket}
\ifx\pdfoutput\undefined
\usepackage[dvipdfmx]{graphicx}
\usepackage[dvipdfmx]{hyperref}
\usepackage[dvipdfmx]{color}
\else
\usepackage{graphicx}
\usepackage{hyperref}
\usepackage{color}
\fi

\begin{document}
\let\emph\textit

\title{
  Ferrimagnetically ordered states in the Hubbard model
  on the hexagonal golden-mean tiling 
}

\author{Akihisa Koga}

\affiliation{
  Department of Physics, Tokyo Institute of Technology,
  Meguro, Tokyo 152-8551, Japan
}

\author{Sam Coates}

\affiliation{
  Department of Materials Science and Technology,
  Tokyo University of Science, Katsushika, Tokyo 125-8585, Japan
}

\date{\today}
\begin{abstract}
  We study magnetic properties of the half-filled Hubbard model on the two-dimensional hexagonal golden-mean tiling.
  We find that the vertex model of the tiling is bipartite,
  with a sublattice imbalance of $\sqrt{5}/(6\tau^3)$ (where $\tau$ is the golden mean),
  and that the non-interacting tight-binding model gives macroscopically degenerate states at $E=0$.
  We clarify that each sublattice has specific types of confined states,
  which in turn leads to an interesting spatial pattern in the local magnetizations in the weak coupling regime.
  Furthermore, this allows us to analytically obtain the lower bound on the fraction of the confined states
  as $(\tau+9)/(6\tau^6)\sim 0.0986$, which is conjectured to be the exact fraction.
  These results imply that a ferrimagnetically ordered state is realized even in the weak coupling limit.
  The introduction of the Coulomb interaction lifts the macroscopic degeneracy at the Fermi level,
  and induces finite staggered magnetization as well as uniform magnetization.
  Likewise, the spatial distribution of the magnetizations continuously changes with increasing interaction strength.
  The crossover behavior in the magnetically ordered states is also addressed
  in terms of the perpendicular space analysis.
\end{abstract}
\maketitle

\section{Introduction}
Quasicrystals and their related phenomena have been of much interest
since the first discovery of the quasicrystalline phase of Al-Mn~\cite{Shechtman}.
Recently, intriguing low-temperature properties originating from electron correlations have been observed
in quasicrystalline and approximant intermetallic phases, including quantum criticality ~\cite{Ishimasa_2011,Deguchi_2012},
heavy fermion behaviour ~\cite{Ishimasa_2011,Deguchi_2012}, and superconductivity ~\cite{Kamiya_2018}.
Particular attention has been paid to magnetic properties in quasicrystals, with the majority of compounds showing spin glass-like states ~\cite{Shechtman,Kimura_1986,SG1,SG2,Chen_1986,SG1,Tsai_1988,AlMnGe,Hattori_1995,Charrier_1997,Sato_1998,Sato_2001}. However, long-range ordered states have been reported in approximant systems ~\cite{Tamura_2010}, and, recently, even in quasicrystals ~\cite{Tamura}.
These novel observations rightfully stimulate further theoretical investigations on electron correlations
inherent in quasicrystalline matter~\cite{Takemori,Takemura,Shinzaki,Hauck}.
A fundamental question is how a quasiperiodic structure affects physical properties in its ordered state. Quasiperiodic tilings provide an exemplar playground for exploring the theoretical answers to this question: investigating,
for example, superconducting and excitonic insulating states ~\cite{Sakai_2017,Sakai_2019,Inayoshi_2020,Cao}.
Indeed, it has been clarified that the effect of the quasiperiodic structure appears in the bulk quantities
in addition to local quantities~\cite{Takemori_2020}. 

To this end, magnetically ordered states on tilings have been well studied ~\cite{Wessel_2003,Jagannathan_2007,Jagannathan_Schulz_1997,Koga_Tsunetsugu,Koga_AB,Koga_dodeca,SakaiKoga},
with interesting magnetic properties reported under the Hubbard models on the Penrose~\cite{Jagannathan_2007,Koga_Tsunetsugu},
Ammann-Beenker~\cite{Jagannathan_Schulz_1997,Koga_AB},
and Socolar dodecagonal tilings~\cite{Koga_dodeca}. Here, no uniform magnetization appears in the thermodynamic limit, and
antiferromagnetically ordered states are always realized when the Coulomb interaction is finite. 
Likewise, in the weak coupling regime, spontaneous magnetization strongly depends on
the distribution of the confined states, while it depends on the local environment in the strong coupling regime.
These results are commonly found in the toy models on these three quasiperiodic tilings.
Therefore, for broader analysis, and the potential for discovering novel properties, it is desirable to examine the magnetic properties of other quasiperiodic tilings.

\begin{figure}[htb]
 \includegraphics[width=\linewidth]{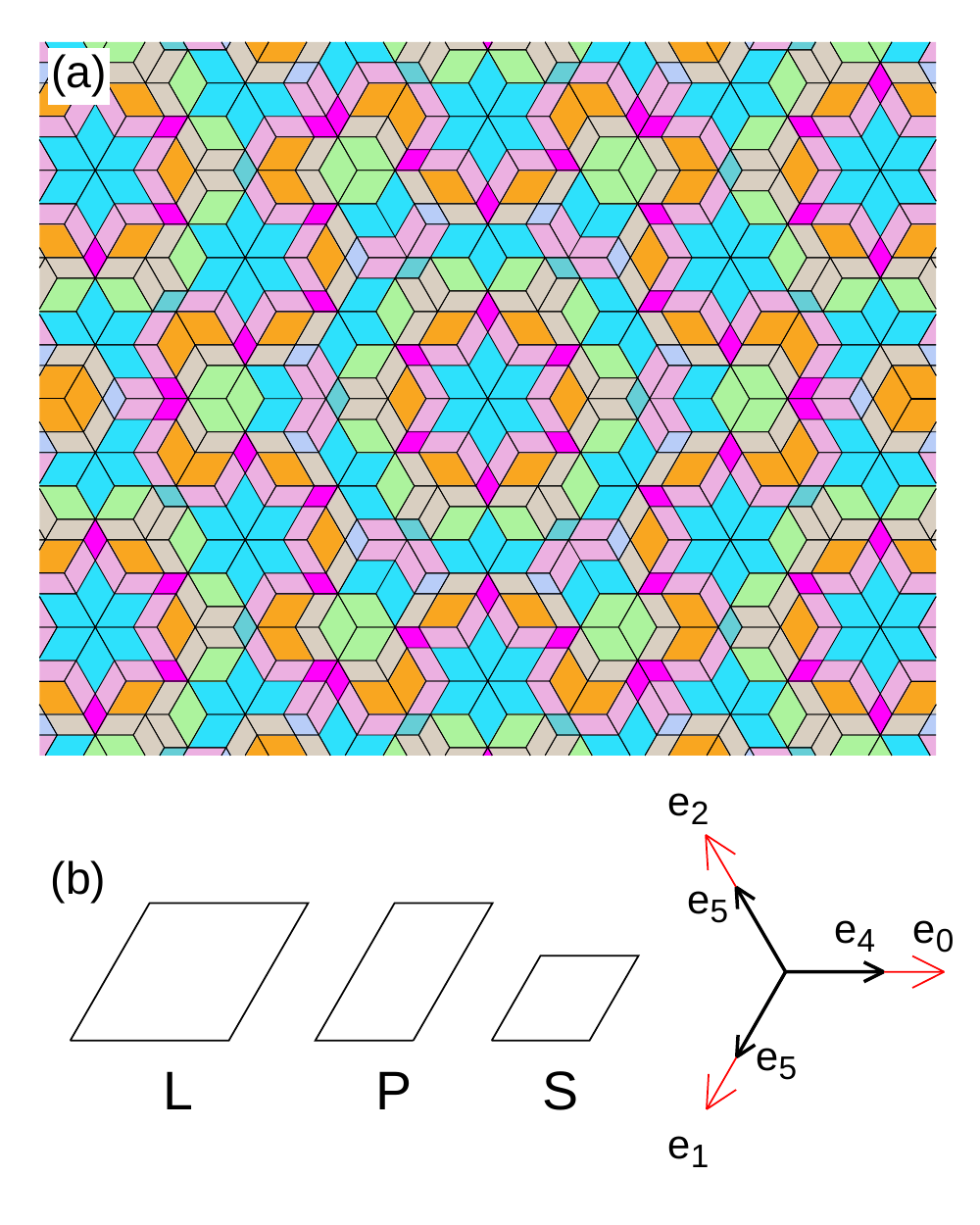}
 \caption{
   (a) Hexagonal golden-mean tiling.
   (b) Large rhombus, parallelogram, and small rhombus.
   ${\bf e}_0, \cdots, {\bf e}_5$ are projection of
   the fundamental translation vectors in six dimensions,
   ${\bf n}=(1,0,0,0,0,0), \cdots, (0,0,0,0,0,1)$.
 }
 \label{lattice}
\end{figure}

In our previous paper~\cite{Sam}, we defined the quasiperiodic  hexagonal golden-mean tiling. The tiling is composed of rhombuses and parallelograms, the vertex model is bipartite, and the system has six-fold rotational symmetry, with a portion shown in Fig.~\ref{lattice}(a). With our definition it is now possible to explore the wider properties of the tiling -- with the aim of understanding and showcasing its place within the quasiperiodic tiling family. In this paper, we examine the tiling structure in detail, with a view to a discussion on its magnetic properties in the half-filled Hubbard model. First, in Sec.~\ref{sec:tiling}, we explain the hexagonal golden-mean tiling in detail, clarifying the existence of a sublattice imbalance which is distinct from that of the Penrose, Ammann-Beenker, and Socolar dodecagonal tilings ~\cite{Koga_Tsunetsugu,Koga_AB,Koga_dodeca}. In Sec.~\ref{sec:model}, we introduce the half-filled Hubbard model on the hexagonal tiling. Then, we study the confined states with $E=0$ in Sec.~\ref{sec:conf}, which should play an important role for magnetic properties in the weak coupling limit. We find that the confined state properties are also distinct from the above cases, and obtain the fraction of the confined states.
By means of the real-space Hartree approximations,
we then clarify how a ferrimagnetically ordered state is realized in the Hubbard model in Sec.~\ref{sec:result}.
Finally, crossover behavior in the ordered state is addressed
by mapping the spatial distribution of the magnetization to the perpendicular space. 



\section{Hexagonal golden-mean tiling}\label{sec:tiling}
\begin{figure*}
    \includegraphics[width=\textwidth]{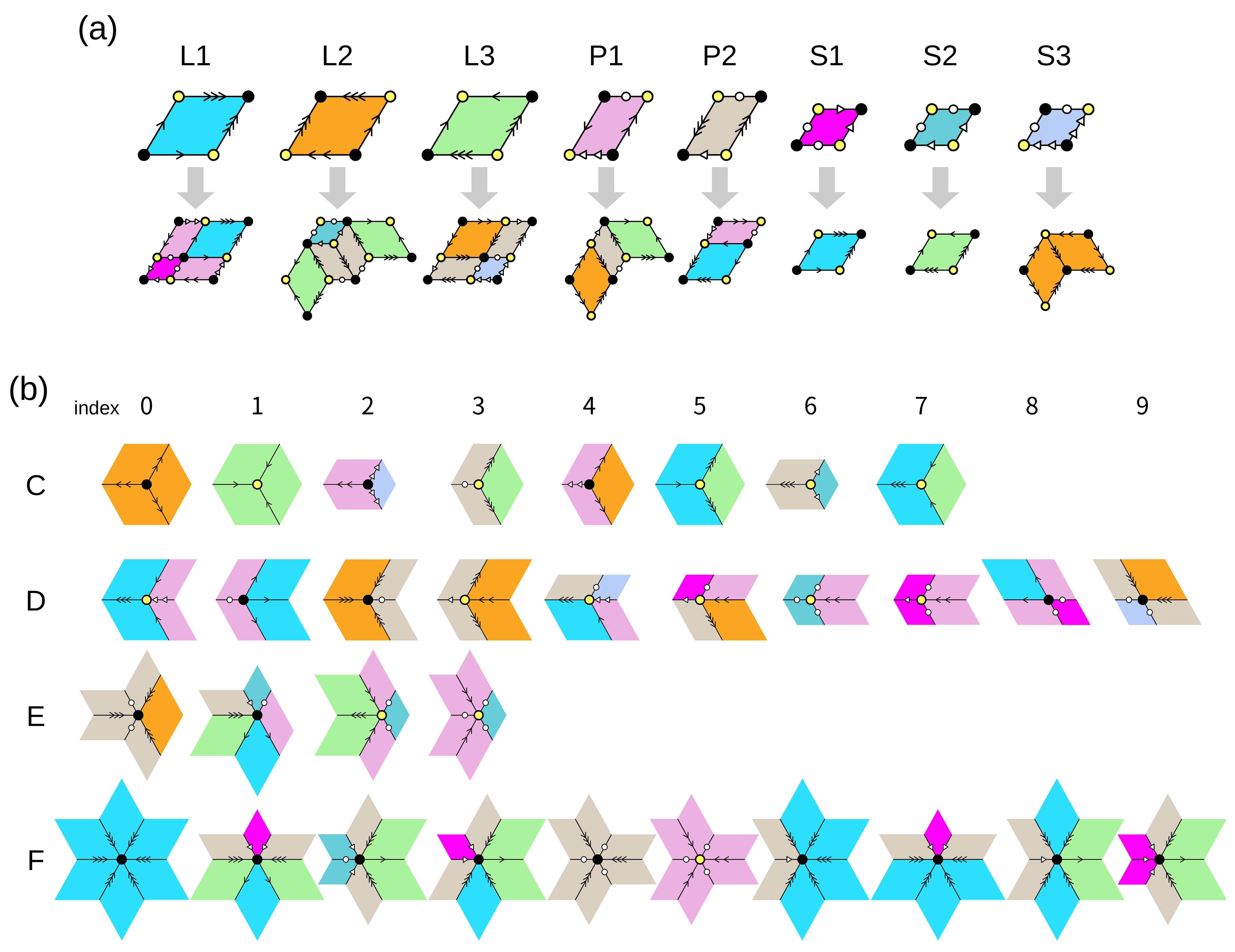}
    \caption{
      (a) Inflation-deflation rule for each rhombus and parallelogram
      in the hexagonal golden-mean tiling~\cite{Sam}.
      Solid and open circles at the corners indicate the ``spin'' (see text).
      (b) Thirty-two types of vertices in the hexagonal golden-mean tiling.
      Solid (open) circles at the vertices represent the sublattice A (B).
    }
    \label{vertices}
  \end{figure*}

The hexagonal golden-mean tiling is composed of large rhombuses (L), parallelograms (P),
and small rhombuses (S),
which are schematically shown in Fig.~\ref{lattice}(b), where the ratio of the longer and smaller lengths of their edges is the golden ratio $\tau = (1+\sqrt{5})/2$ ~\cite{Sam}. The tiling can be generated using grid-dualization or high-dimension projection, but here, we make use of the substitution rules
to generate its structure,
since it is straightforward to obtain the exact fractions for various diagrams in the thermodynamic limit.
In this case, the 3 basic tiles are further characterized by eight types of substitutions. The rules for this extended tile-set are schematically shown in Fig.~\ref{vertices}(a), where we define three large rhombuses (L1, L2, L3),
two parallelograms (P1, P2), 
and three small rhombuses (S1, S2, S3). In this schematic we also decorate the tiles with arrows, open triangles, and open circles on their edges, to demonstrate and satisfy the matching rules we have previously defined ~\cite{Sam}. The fractions of the tiles in the infinite tiling are given as:
\begin{eqnarray}
  f_{\rm L1}&=&\frac{1}{2\tau^2}\sim 0.191,\label{eq:L1}\\
  f_{\rm L2}&=&\frac{1}{4\tau^2}\sim 0.095,\\
  f_{\rm L3}&=&\frac{1}{4\tau^2}\sim 0.095,\\
  f_{\rm P1}&=&\frac{1}{\tau^3}\sim 0.236,\\
  f_{\rm P2}&=&\frac{1}{\tau^3}\sim 0.236,\\
  f_{\rm S1}&=&\frac{1}{2\tau^4}\sim 0.073,\\
  f_{\rm S2}&=&\frac{1}{4\tau^4}\sim 0.036,\\
  f_{\rm S3}&=&\frac{1}{4\tau^4}\sim 0.036.\label{eq:S3}
\end{eqnarray}

The tiling contains thirty-two distinct types of vertices, which are classified into four groups specified by their coordination numbers $Z$, with $Z=3, 4, 5$, and $6$, which we label as C, D, E, and F vertices, respectively.
Each vertex is specified by its group and an index $i$,
as explicitly shown in Fig.~\ref{vertices}(b). The fractions of the vertices can be obtained by examining the deflation rule~\cite{Sam}, with these values shown in Table~\ref{table}.
We note that the average of the coordination number is 4,
which is the same for the vertex model
on the Penrose and Ammann-Beenker tilings, and square lattice.
Therefore, one may expect that magnetic properties in the hexagonal golden-mean tiling
are similar to those on the above lattices.

We note that there are some vertices with local rotational symmetry which exhibit interesting substitution behaviour. For instance, the F$_0$ vertex is located at the centre of six adjacent L1 rhombuses, and thereby
has local sixfold rotational symmetry. Under one substitution of its surrounding tiles [see Fig.~\ref{vertices}(a)],
we find that the F$_0$ vertex remains unchanged, which is behaviour unique to this position. Likewise, there are four vertices with
local threefold rotational symmetry, such as
C$_0$, C$_1$, F$_4$, and F$_5$.
According to the deflation rule, these vertices are changed in a cyclical manner as
${\rm C}_1 \rightarrow {\rm C}_0 \rightarrow {\rm F}_4 \rightarrow {\rm F}_5 \rightarrow {\rm C}_1 \rightarrow\cdots$.  Both of these properties will be important when examining the number of confined states in the tiling.

\begin{center}
  \begin{table}
    \caption{Fractions of C$_i$, D$_i$, E$_i$, and F$_i$ vertices
      in the hexagonal golden-mean tiling.
      The asterisks indicate that the vertices are located in
      the B sublattice (see text). 
    }
    \begin{tabular}{c|cccccccccc}
      \hline
      \hline
      &0&1&2&3&4&5&6&7&8&9\\
      \hline
      C&$\displaystyle\frac{1}{12\tau^6}$&$\displaystyle\frac{1}{12\tau^4}^*$&
      $\displaystyle\frac{1}{4\tau^4}$&$\displaystyle\frac{\sqrt{5}}{4\tau^4}^*$&
      $\displaystyle\frac{\sqrt{5}}{4\tau^4}$&$\displaystyle\frac{1}{4\tau^6}^*$&
      $\displaystyle\frac{1}{4\tau^4}^*$&$\displaystyle\frac{1}{4\tau^3}^*$&\\
      D&$\displaystyle\frac{\sqrt{5}}{4\tau^4}^*$&$\displaystyle\frac{1}{4\tau^6}$&
      $\displaystyle\frac{1}{4\tau^6}$&$\displaystyle\frac{1}{4\tau^6}^*$&
      $\displaystyle\frac{1}{2\tau^4}^*$&$\displaystyle\frac{3}{2\tau^5}^*$&
      $\displaystyle\frac{1}{4\tau^8}^*$&$\displaystyle\frac{1}{4\tau^8}^*$&
      $\displaystyle\frac{1}{2\tau^4}$&$\displaystyle\frac{1}{4\tau^4}$\\
      E&$\displaystyle\frac{\sqrt{5}}{4\tau^6}$&$\displaystyle\frac{\sqrt{5}}{2\tau^6}$&
      $\displaystyle\frac{1}{4\tau^6}^*$&$\displaystyle\frac{\sqrt{5}}{4\tau^8}^*$&\\
      F&$\displaystyle\frac{1}{4\tau^6}$&$\displaystyle\frac{1}{2\tau^6}$&
      $\displaystyle\frac{1}{4\tau^8}$&$\displaystyle\frac{\sqrt{5}}{2\tau^8}$&
      $\displaystyle\frac{1}{12\tau^8}$&$\displaystyle\frac{1}{12\tau^{10}}^*$&
      $\displaystyle\frac{\sqrt{5}}{4\tau^8}$&$\displaystyle\frac{1}{2\tau^8}$&
      $\displaystyle\frac{1}{4\tau^{10}}$&$\displaystyle\frac{1}{4\tau^8}$
      \\
      \hline
      \hline
    \end{tabular}
    \label{table}
  \end{table}
\end{center}

Now we consider the sublattice structure in the hexagonal golden-mean tiling,
which is important for discussing magnetic properties in the Hubbard model.
To do so,
we first must introduce a ``spin'' for the tiles
as (L1)$_\sigma$, (L2)$_\sigma$, $\cdots$, (S3)$_\sigma$
with the spin $\sigma(=\uparrow, \downarrow)$,
which uniquely specifies spins at the corner sites.
To accurately assign spins to each tile considering their vertex environments,
we have considered the substitution rules under the vertex scheme.
For example, as discussed above, when one applies the deflation operation to the tiling,
${\rm F}_{0}$ vertices remain as ${\rm F}_{0}$ vertices.
Therefore, the spin configuration for the tile (L1)$_\sigma$ and its deflation rule can be defined so that
the spins at the corner sites with acute angles are not changed under the deflation operation.
Then, the deflation rule for the tile (L1)$_\sigma$ is described as
${\rm (L1)}_\sigma \rightarrow {\rm (L1)}_\sigma+ 2{\rm (P1)}_\sigma+{\rm (S1)}_\sigma$
when the spin configurations for the tiles (P1)$_\sigma$ and (S1)$_\sigma$ are defined,
as shown in Fig.~\ref{vertices}(a). 
In the tiling, the tile P1 always appears next to the tile L1, and
we consider the substitution rule for these adjacent tiles.
This gives the deflation rule for the tile (P1)$_\sigma$ as
${\rm (P1)}_\sigma\rightarrow 1/2 {\rm (L2)}_\sigma+1/2{\rm (L3)}_\sigma+{\rm (P2)}_\sigma$,
when the spin configurations for the tiles (L2)$_\sigma$, (L3)$_\sigma$, and (P2)$_\sigma$
is defined, as shown in Fig.~\ref{vertices}(a).
Continuing in this fashion for the other tiles and
taking into account the matching rules for the other tiles in the tiling,
we therefore define the spin-dependent tiles and their deflation rules in Fig.~\ref{vertices}(a), where
solid circles represent the spin $\sigma$ and open circles spin $\bar{\sigma}$.

To demonstrate, the number of tiles are changed under the deflation operation as
${\bf v}^{n+1}_\sigma=M {\bf v}^n_\sigma$,
where $({\bf v}^n_\sigma)^t=(N^n_{{\rm (L1)}_\sigma},N^n_{{\rm (L2)}_\sigma},\cdots,N^n_{{\rm (S3)}_\sigma})$,
$N^n_{{\rm Q}_\sigma}$ is the number of tiles Q$_\sigma$ at iteration $n$, and
\begin{equation}
M=\left(
\begin{array}{cccccccc}
1 & 0 & 0 & 0 & 1 & 1 & 0 & 0\\
0 & 0 & 1 & \frac{1}{2}&0 & 0 & 0 & 1\\
0 & 1 & 0 & \frac{1}{2}&0 & 0 & 1 & 0\\
2 & 0 & 0 & 0 & 1 & 0 & 0 & 0\\
0 & 2 & 2 & 1 & 0 & 0 & 0 & 0\\
1 & 0 & 0 & 0 & 0 & 0 & 0 & 0\\
0 & 1 & 0 & 0 & 0 & 0 & 0 & 0\\
0 & 0 & 1 & 0 & 0 & 0 & 0 & 0
\end{array}
\right).
\end{equation}
This transformation matrix for the number of spin-dependent tiles
is diagonal with respect to the spin, implying that tiles with only one of the spins will appear in the thermodynamic limit.
This property is distinct from that of the other two-dimensional quasiperiodic tilings
such as the Penrose~\cite{Koga_Tsunetsugu},
Ammann-Beenker~\cite{Koga_AB}, and Socolar dodecagonal tilings~\cite{Koga_dodeca},
where spin-dependent tiles appear equally in the thermodynamic limit.
Immediately, we find a sublattice imbalance in the vertices in the hexagonal golden-mean tiling.
The thirty-two types of vertices are uniquely classified into
two sublattices, which are shown as solid or open circles in Fig.~\ref{vertices}(b) as a result of our spin-decoration. For convenience, we refer to the sublattice which includes the F$_0$ vertices (solid) as sublattice A, and the other (open) as sublattice B.
The fractions of the sublattices A and B are obtained by summing the relevant terms of Table~\ref{table} as
$f_A=1/2-\sqrt{5}/(12\tau^3)$ and $f_B=1/2+\sqrt{5}/(12\tau^3)$.
Then, the sublattice imbalance is given
as
\begin{eqnarray}
  \big|f_A-f_B\big|=\frac{\sqrt{5}}{6\tau^{3}}\sim 0.088.\label{sub}
\end{eqnarray}
We note that the sublattice imbalance originates
inherently from the structure of the hexagonal golden-mean tiling.
Therefore, the sublattice imbalance is represented by an irrational number, which is distinct from the trivial case in bipartite decorated lattices;
for example, the Lieb lattice with $\big|f_A-f_B\big|=1/3$.
In the following, we omit the spin index in the tiles and vertices
to discuss magnetic properties 
in the correlated electron system on the hexagonal golden-mean tiling.

\section{Model and Hamiltonian}\label{sec:model}
We study the Hubbard model on the hexagonal golden-mean tiling,
which is given by the following Hamiltonian:
\begin{eqnarray}
  H&=&-t\sum_{(ij),\sigma}\left(c_{i\sigma}^\dag c_{j\sigma}+h.c.\right)
  +U\sum_i n_{i\uparrow}n_{i\downarrow},\label{H}
\end{eqnarray}
where $c_{i\sigma} (c_{i\sigma}^\dag)$ annihilates (creates) an electron
with spin $\sigma(=\uparrow, \downarrow)$ at the $i$th site and
$n_{i\sigma}=c_{i\sigma}^\dag c_{i\sigma}$.
$t$ denotes the nearest neighbor transfer integral
and $U$ denotes the onsite Coulomb interaction.
The chemical potential is always $\mu=U/2$ when the electron
density is fixed to be half filling.

To discuss magnetic properties in the Hubbard model,
we make use of the real-space mean-field theory.
This method has an advantage in treating large clusters,
which is crucial to clarifying magnetic properties in the system with a quasiperiodic tiling.
Here, we introduce the site-dependent mean-field $\langle n_{i\sigma}\rangle$.
The mean-field Hamiltonian is then given as
\begin{eqnarray}
  H^{MF}&=&-t\sum_{(ij),\sigma}\left(c_{i\sigma}^\dag c_{j\sigma}+h.c.\right)
  +U\sum_{i,\sigma} n_{i\sigma}\langle n_{i\bar{\sigma}}\rangle.\label{H}
\end{eqnarray}
For given values of $\langle n_{i\sigma}\rangle$,
we numerically diagonalize the Hamiltonian $H^{MF}$, update
$\langle n_{i\sigma}\rangle$, and repeat this procedure until the result
converges.
The uniform and staggered magnetizations $m^\pm$ are given as
\begin{eqnarray}
  m^\pm &=&f_A m_A \pm f_B m_B,\\
  m_{\alpha}&=&\frac{1}{N_{\alpha}}\sum_{i\in \alpha} m_i,\\
  m_i&=&\frac{1}{2}\left( \langle n_{i\uparrow} \rangle
  -\langle n_{i\downarrow} \rangle\right),
\end{eqnarray}
where $N_\alpha$ ($m_\alpha$) is the number of sites
(average of the magnetization) in the sublattice $\alpha(=A, B)$,
and $m_i$ is the magnetization at the $i$th site.
According to Lieb's theorem~\cite{Lieb},
the half-filled Hubbard model on the bipartite lattice
has a ground state with
total spin $S_{tot} =\frac{1}{2}|N_A-N_B|$.
In fact, it has been clarified that
the magnetically ordered states with finite total spin
are realized in some periodic Hubbard systems~\cite{Lieb,Mielke,Tasaki,Kusakabe,Noda}.
As we have discussed, the vertex model on the hexagonal golden-mean tiling has
a sublattice imbalance, so that
the magnetically ordered state is realized with
a total spin $\sqrt{5}/(12\tau^3)\times N$,
where $N(=N_A+N_B)$ is the total number of sites.

\begin{figure}[htb]
 \includegraphics[width=\linewidth]{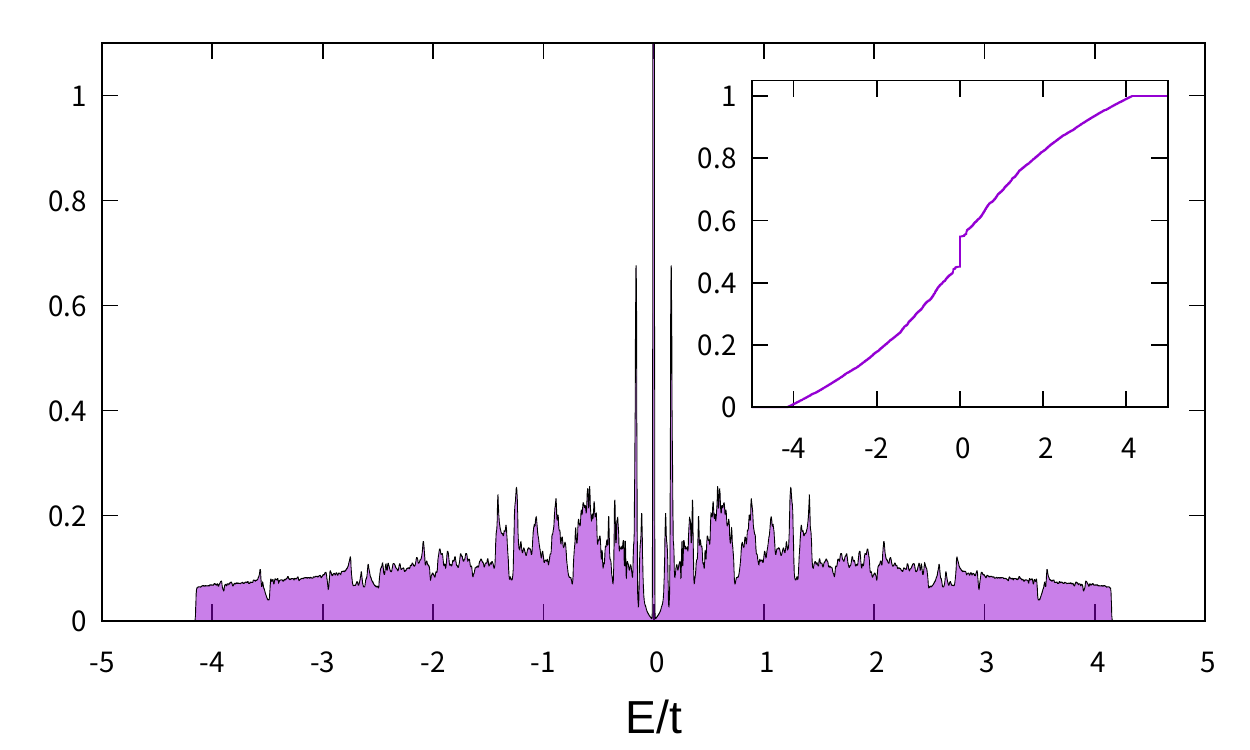}
 \caption{
   Density of states of the tightbinding model
   on the hexagonal golden-mean tiling with $N=1\,172\,071$.
   The inset shows the integrated density of states. 
 }
 \label{fig:dos}
\end{figure}

When the system is noninteracting ($U=0$),
the model Hamiltonian is reduced to the tightbinding model.
The density of states for the system with $N=1\,172\,071$
is shown in Fig.~\ref{fig:dos}, which is symmetric since the system is bipartite.
We also find the delta-function peak at $E=0$,
which implies the existence of macroscopically degenerate states.
These confined states should be important for magnetic properties
in the weak coupling limit, and are features common to
bipartite quasiperiodic tilings.
In addition, we find sharp peaks at $E/t\sim \pm 0.16$,
but, these have small widths, so that the corresponding eigenstates are not
strictly localized in a certain region
in contrast to the confined states with $E=0$.


\section{Confined states}\label{sec:conf}
\begin{figure}[htb]
 \includegraphics[width=\linewidth]{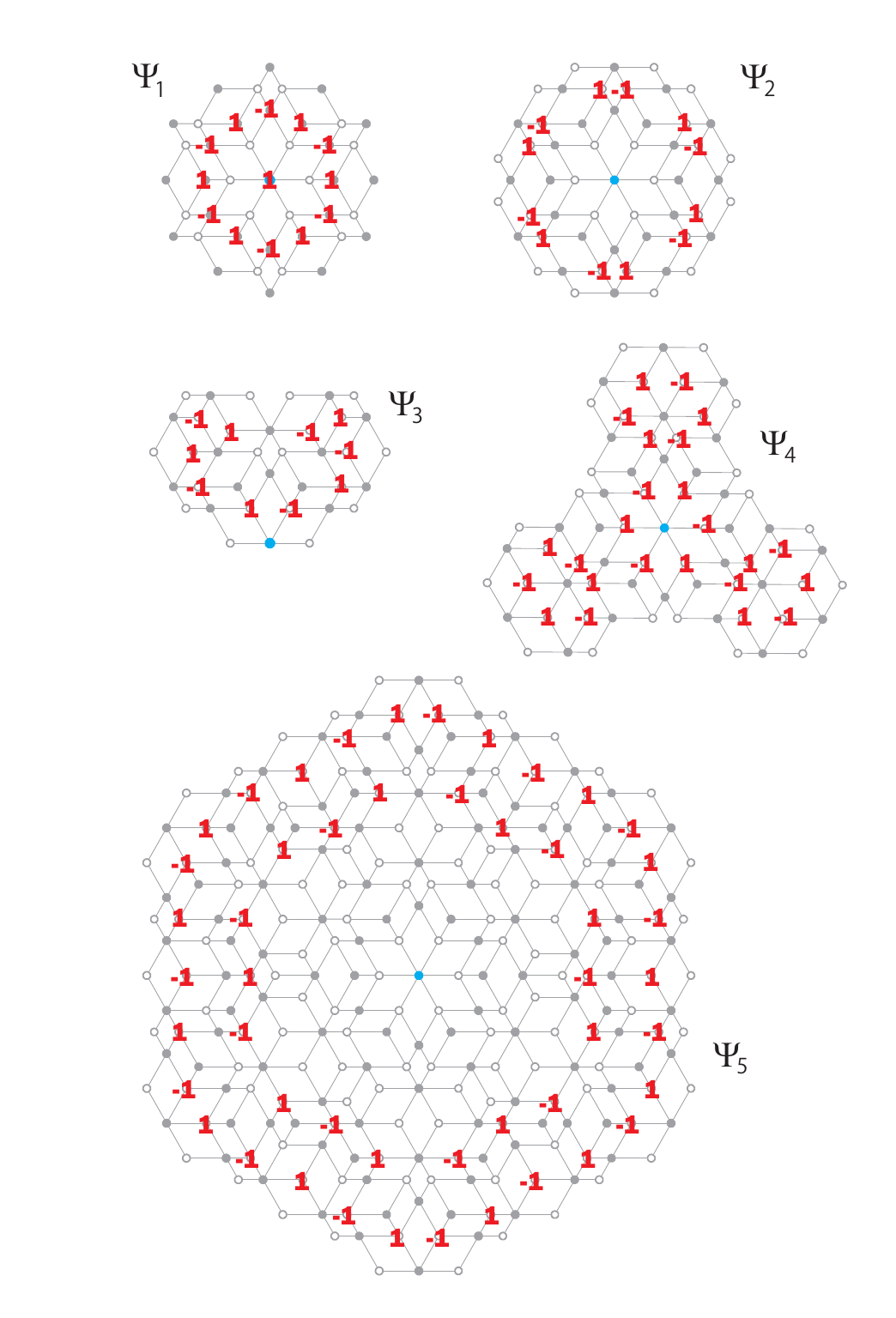}
 \caption{
   Five confined states around the F$_0$ vertex (blue points).
   Solid (open) circles represent the sublattice A (B).
   The number at the vertices represent the amplitudes of the confined states.
 }
 \label{confX}
\end{figure}

Here, we study the confined states in detail; since the confined states with $E=0$ are macroscopically degenerate,
we can choose a simple form by considering their linear combinations,
as discussed in several papers~\cite{KohmotoSutherland,Arai,Koga_Tsunetsugu}.
In the hexagonal golden-mean tiling, 
infinite types of confined states should appear,
similar to the Ammann-Beenker and Socolar dodecagonal tilings~\cite{Koga_AB,Koga_dodeca}.
Some simple examples of confined states around the F$_0$  vertices are
explicitly shown in Fig.~\ref{confX}, labelled as $\Psi_1$, ..., $\Psi_5$.
We find two confined states, $\Psi_1$ and $\Psi_2$,
in the smallest region with sixfold rotational symmetry.
The amplitudes in $\Psi_1$ appear at the F$_0$ vertex and
next-nearest neighbor C$_6$ and D$_8$ vertices in the sublattice A.
By contrast, the amplitudes in $\Psi_2$
appear at the D$_5$ vertices in the sublattice B.
Since the confined states have amplitudes in both sublattices,
we can say that the introduction of the Coulomb interaction
lifts the degeneracy at $E=0$,
stabilizing the magnetically ordered state with
a finite staggered magnetization.
We note that these confined states do not necessarily exist
around all the F$_0$ vertices.
In fact, each fraction $(\Psi_1, \Psi_2)$ is given as $1/(4\tau^8)$ (smaller than $f_{{\rm F}_0}$).
The fractions of the other examples of confined states $\Psi_3$, $\Psi_4$, and $\Psi_5$ are
given by $5/(4\tau^{10})$, $2/(4\tau^{10})$, and $1/(4\tau^{12})$,
by taking each local symmetry into account.

\begin{figure}[htb]
 \includegraphics[width=\linewidth]{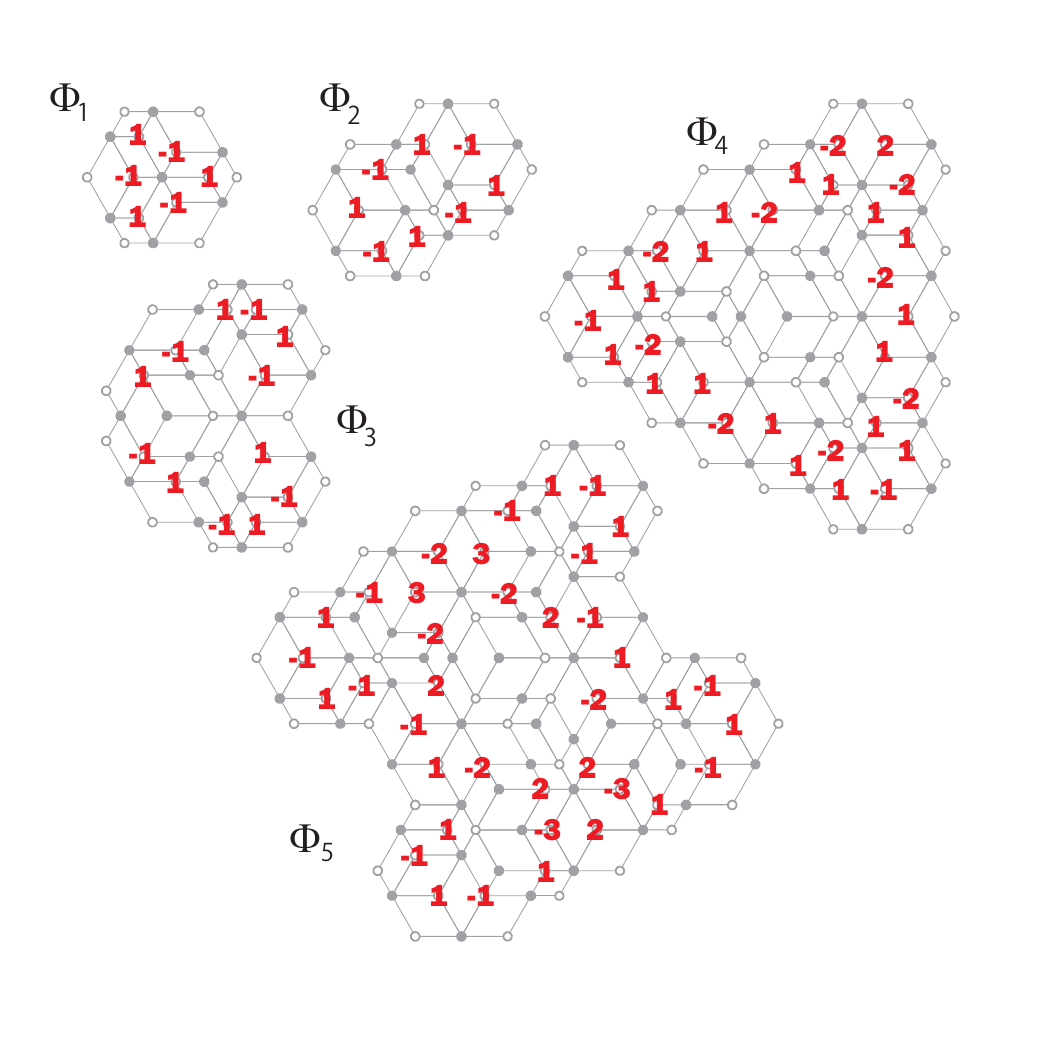}
 \caption{
   Five confined states. Solid (open) circles represent the sublattice A (B).
   The number at the vertices represent the amplitudes of the confined states.
 }
 \label{confT}
\end{figure}
Away from the F$_0$ vertices, Figure~\ref{confT} shows five examples of confined states around different vertices.
Wave functions $\Phi_1$ and $\Phi_4$ are located around
the F$_4$ and C$_0$ vertices, respectively, which exhibit threefold rotational symmetry, while $\Phi_2$, $\Phi_3$, and $\Phi_5$ are located
at vertices which locally have no rotational symmetry.
As seen in Figs.~\ref{confX} and \ref{confT},
many kinds of confined states are expected in the thermodynamic limit.
Therefore, we can not count the number of confined states
systematically, in contrast to the Ammann-Beenker tiling, for example.

We wish to note that the confined states having amplitudes in the sublattice A
should be restricted to $\Psi_1$, as shown in Figs.~\ref{confX} and \ref{confT}.
We can not prove this analytically, but rather confirm it numerically in a cluster with $N=448\,213$,
which will be shown later.
This conjecture immediately gives us the lower bound of
the fraction of the confined states,
by considering magnetic properties in the weak coupling limit.
In the limit, the confined states play an essential role
for magnetic properties, and the uniform magnetization is given as
$m^+=(f^C_{A}-f^C_{B})/2$,
where $f^C_{A}$ ($f^C_B$) is the fraction of the confined states
in the sublattice A (B).
Namely, as before, $f^C_A=1/(4\tau^8)$.
The uniform magnetization is directly related to
the sublattice imbalance eq.~(\ref{sub}).
As such, we obtain the fraction of the confined states as
\begin{eqnarray}
  f^C=f^C_A+f^C_B=\frac{\tau+9}{6\tau^6}\sim 0.0986.
\end{eqnarray}
This value is conjectured to be an exact fraction, and is consistent with the numerical results
$f^C=0.098168$ for a cluster with $N=835\,393$,
where the eigenstates with $E=0$ around the edge are excluded.

\section{Magnetic properties}\label{sec:result}
\begin{figure}[htb]
 \includegraphics[width=\linewidth]{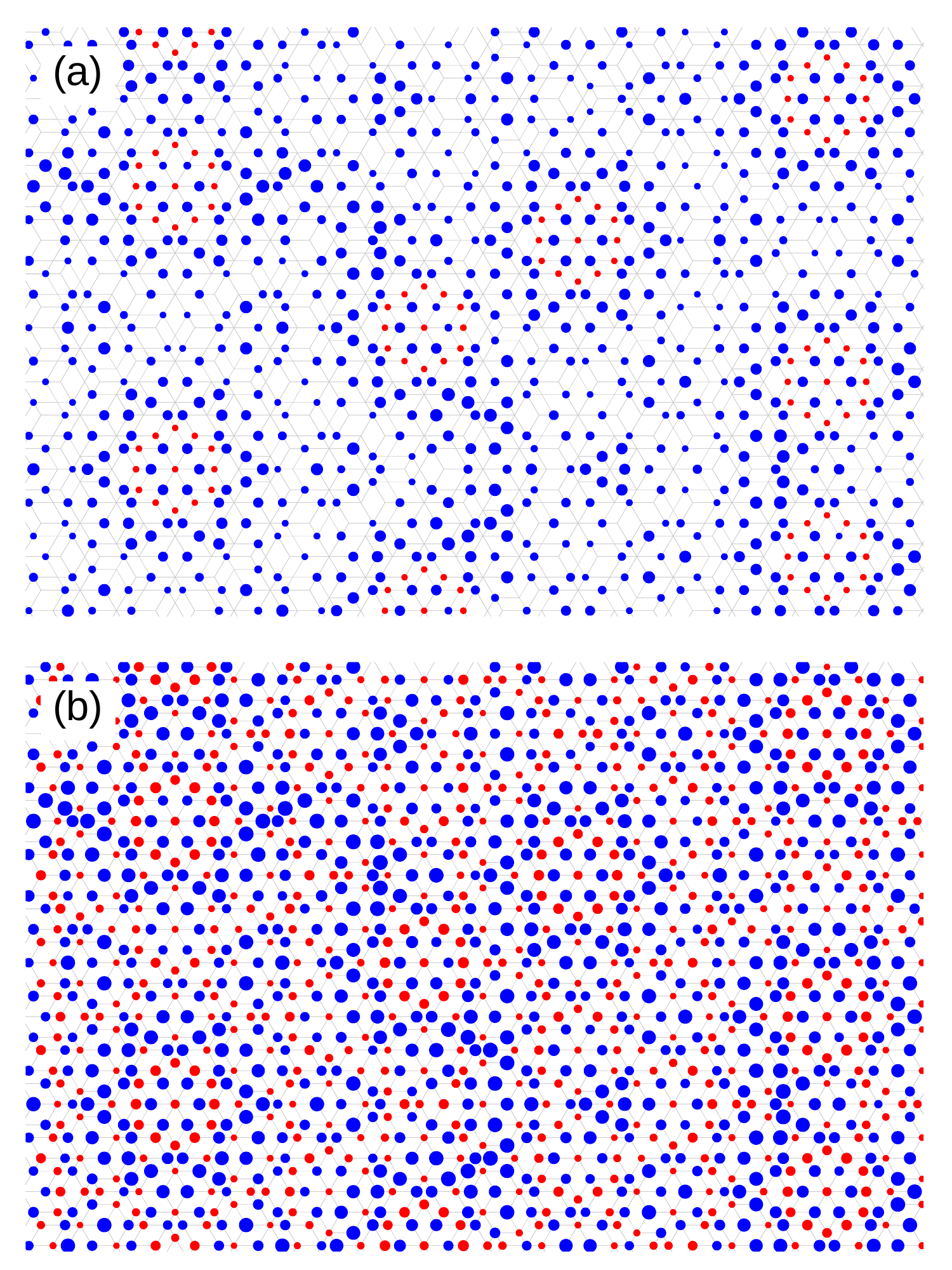}
 \caption{
   Spatial pattern for the staggered magnetizations in the Hubbard model
   on the hexagonal golden-mean tiling when
   $U/t=1.0\times 10^{-6}$ (essentially the same as $U=0$) and $1$.
   The area of the circles represents the magnitude of the local magnetization.
 }
 \label{fig:local}
\end{figure}

In the following, we discuss magnetic properties
in the Hubbard model on the hexagonal golden-mean tiling.
We mainly treat the system with $N=448\,213$ by means of
the real-space mean-field approximations.
When the system is non-interacting,
the macroscopically degenerate states in the density of states
appear at the Fermi level,
as shown in Fig.~\ref{fig:dos}.
The introduction of the interaction lifts the degeneracy,
stabilizing the magnetically ordered state.
The magnetization profile for the case with $U/t=1.0\times 10^{-6}$
is shown in Fig.~\ref{fig:local}(a).
We find finite magnetizations at certain sites, which reflects the spatial distribution of the confined states.
We note that, in the sublattice A,
finite magnetization only appears at the F$_0$ vertices
and their next-nearest-neighbor C$_4$ and D$_8$ vertices, (shown in red).
This is consistent with the fact that, in the sublattice A,
 $\Psi_1$ is the only type of confined state.
Therefore, the magnetization at these vertices take $m_i=1/26$.
In contrast, each site on the sublattice B has
a local magnetization, as shown by the blue circles in Fig.~\ref{fig:local}(a), where the area of the circles represents the magnitude of the local magnetization.
This implies that at each site in the sublattice B,
some confined states have amplitudes.
For this reason, non-magnetic sites only belong to the sublattice A, and
their fraction should be given as $f_{non}=f_A-13f_A^C\sim 0.387$.
Imbalanced magnetic properties in the weak coupling limit are distinct
from those for the quasiperiodic tilings
such as Penrose, Ammann-Beenker, and Socolar dodecagonal tilings~\cite{Koga_Tsunetsugu,Koga_AB,Koga_dodeca}.
Namely, the average of the total uniform and staggered magnetizations
are given as $|m^+|=0.044$ and $|m^-|=0.049$.

\begin{figure}[htb]
 \includegraphics[width=\linewidth]{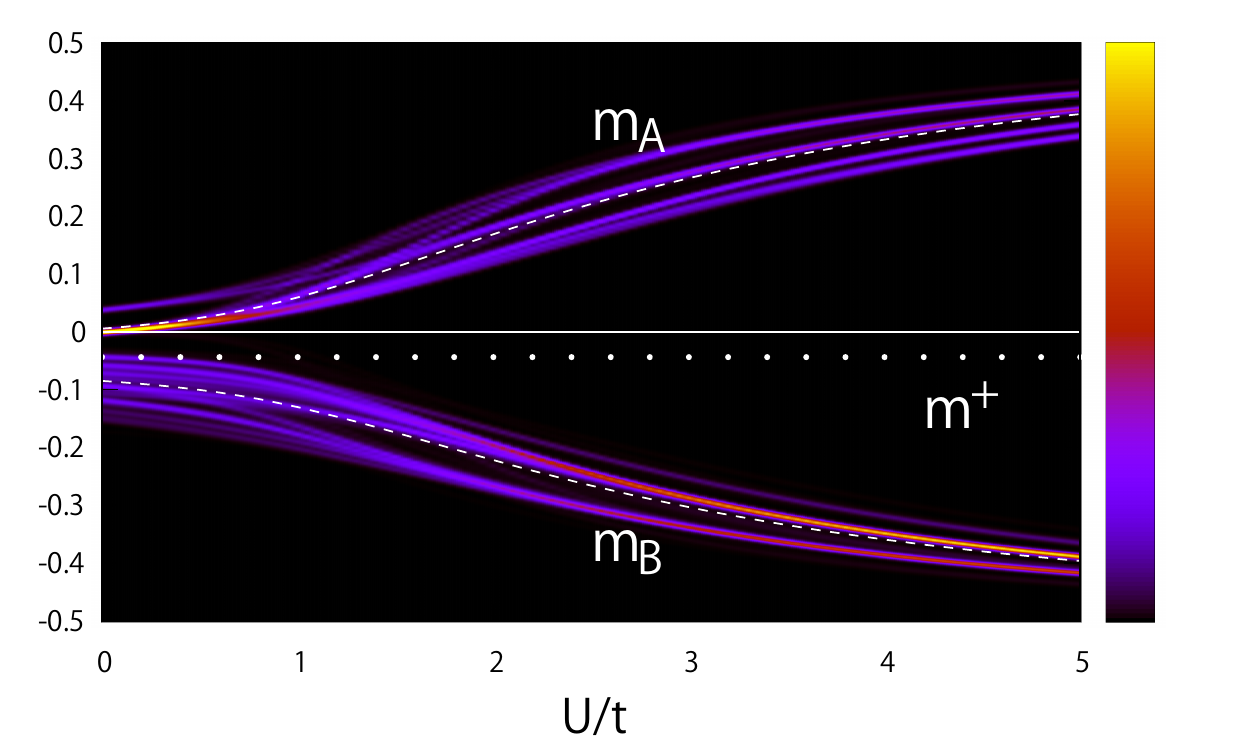}
 \caption{
   Distribution of the local magnetizations 
   as a function of $U/t$ in the system with $N=448\,213$.
   Dashed lines represent the magnetizations in both sublattices
   and the dotted line represents the total uniform magnetization.
 }
 \label{mag}
\end{figure}

When we increase the interaction strength, the absolute value of
the magnetization at each site monotonically increases,
and the non-magnetic sites now have positive magnetizations,
as shown in Fig.~\ref{fig:local}(b).
The distribution of the local magnetization is shown in Fig.~\ref{mag}.
When $U/t\le 1$, the distribution is similar to that in the weak coupling limit $U/t\rightarrow 0$,
where a sharp peak appears in the case $m>0$ (sublattice A),
while a broader structure appears in the case $m<0$ (sublattice B).
In contrast, when $U/t\ge 2$,
distinct behavior appears in the magnetic distribution.
In this case, the absolute values of the local magnetization
should be classified into four groups,
specified by the coordination number of the vertices --
although two of four may be invisible in the case $m<0$.
The crossover between weak and strong coupling regimes occurs
around $U/t\sim 1.5$.
In the strong coupling limit $U/t\rightarrow \infty$,
the Hubbard model is reduced to
the antiferromagnetic Heisenberg model with nearest-neighbor couplings $J=4t^2/U$.
The mean-field ground state is then described by the staggered moment $m_j=\pm 1/2$.
This means that the mean-field approach can not 
correctly describe the reduction of the magnetic moment due to quantum fluctuations.
Therefore, an elaborate method is necessary to precisely clarify magnetic properties in this regime, which is beyond the scope of the present study.
Nevertheless, interesting magnetic properties inherent in the hexagonal golden-mean tiling
can be captured correctly, even in our simple mean-field method;
note that the total uniform magnetization is never changed,
which is consistent with Lieb's theorem~\cite{Lieb}.

\begin{figure}[htb]
 \includegraphics[width=\linewidth]{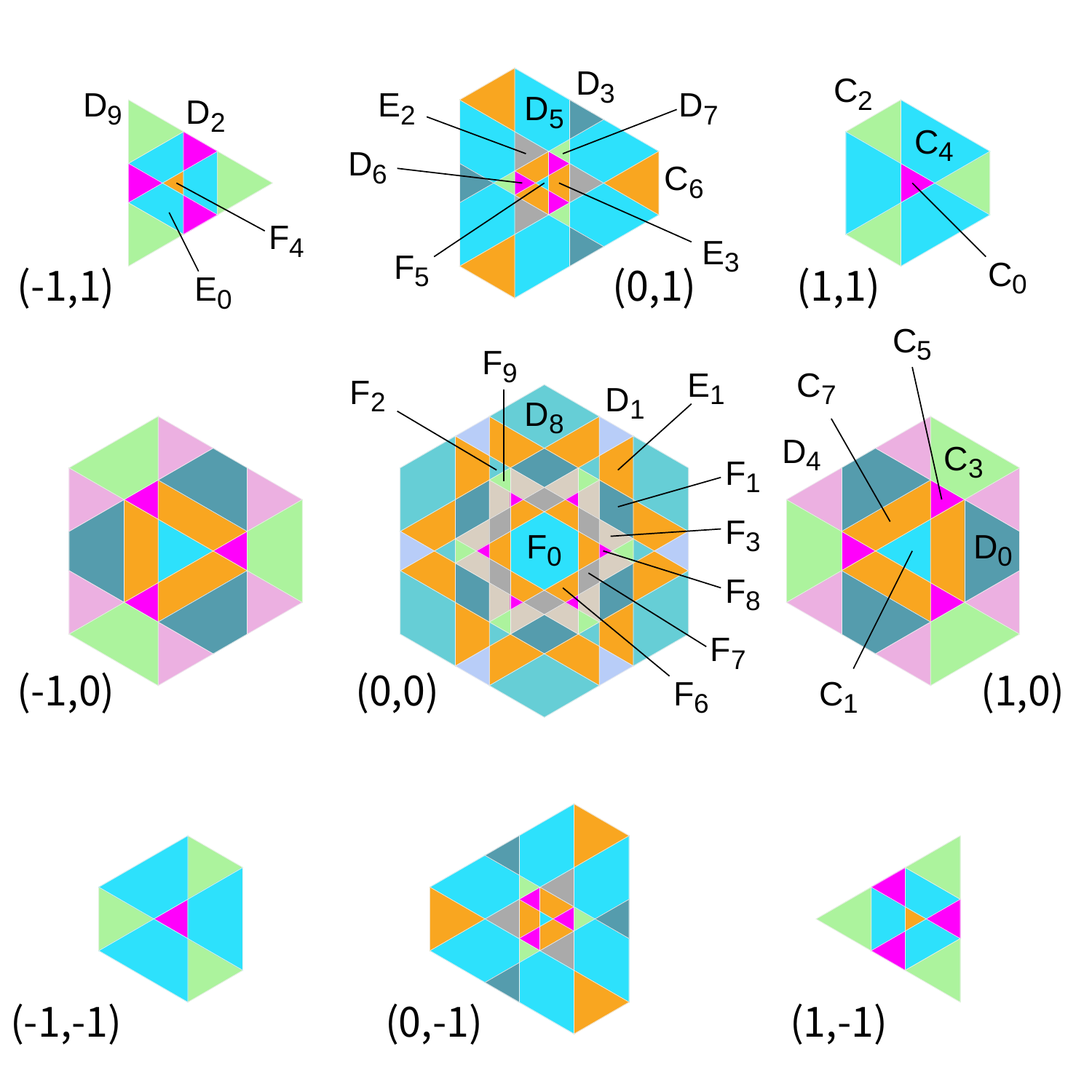}
 \caption{
   Perpendicular spaces ${\bf r}^\perp$ for the hexagonal golden-mean tiling.
   Each area bounded by the solid lines is the region of one of thirty-two types of vertices
   shown in Fig.~\ref{vertices}(b).
 }
 \label{perp}
\end{figure}

Finally, we wish to demonstrate the spatial profile of the magnetizations
characteristic of the hexagonal golden-mean tiling.
To this end, we map it to the perpendicular space, where the positions in perpendicular space have one-to-one correspondence with
the positions in the physical space.
Each site in the tiling is described by a six-dimensional
lattice point $\vec{n}=(n_0, n_1, \cdots, n_5)$, labelled with integers $n_m$, where the lattice is spanned by fundamental translation vectors.
The coordinates of the tiling are the projections onto the two-dimensional
physical space:
\begin{eqnarray}
{\bf r} = (x, y) = \sum_{m=0}^5 n_m{\bf e}_m,
\end{eqnarray}
where ${\bf e}_m$ are the projected vectors and are given as $=\tau(\cos(-2/3\pi m),\sin(-2/3\pi m))$ for $m=0,1,2$ and
${\bf e}_m=(\cos(-2/3\pi m),\sin(-2/3\pi m))$ for $m=3,4,5$, shown in Fig.~\ref{lattice}(b).
We can then project the points onto the four-dimensional perpendicular space (split into two two-dimensional spaces $\tilde{\bf r}$ and $ {\bf r}^\perp$), giving information
specifying the local environment of each site:
\begin{eqnarray}
  \tilde{\bf r}&=&\sum_{m=0}^5n_m\tilde{\bf e}_m,\\
  {\bf r}^\perp&=&\sum_{m=0}^5n_m{\bf e}_m^\perp,
\end{eqnarray}
where $\tilde{\bf e}_m={\bf e}_{m+3}$ for $m=0,1,2$ and
$\tilde{\bf e}_m=-{\bf e}_{m-3}$ for $m=3,4,5$.
${\bf e}^\perp=(1,0)$ for $m=0,1,2$, and ${\bf e}^\perp=(0,1)$ for $m=3,4,5$.
In the hexagonal golden-mean tiling, ${\bf r}^\perp=(x^\perp,y^\perp)$
takes nine values $x^\perp=-1,0,1$ and $y^\perp=-1,0,1$.
In each ${\bf r}^\perp$ plane, the $\tilde{\bf r}$ points densely
cover a certain region.
Moreover, the region in plane ${\bf r}^\perp$ has the same size as the one in the plane $(-{\bf r}^\perp)$.  
Fig.~\ref{perp} shows the perpendicular space for the system.
The plane $(0,0)$ has sixfold rotational symmetry, while
the others have threefold rotational symmetry.
We have previously shown that the thirty-two types of vertices are mapped into specific
regions in certain planes \cite{Sam}.
This implies that the perpendicular space reflects the local environments
for the lattice sites, such that the areas of each vertex region in perpendicular space are proportional to its fraction in parallel space.
As such, we also find that the vertices in the sublattice A (B) are mapped to the planes
with $(0,0)$, $(\pm 1,\pm 1)$ and $(\pm 1,\pm 1)$ [$(0, \pm 1)$ and $(\pm 1,0)$].
This can be explained by the following:
The sublattice index for each vertex is uniquely determined, as discussed above.
Since upon moving from one site to its neighbor only one
of $n_m$'s changes by $\pm 1$,
even (odd) number $(x^\perp+y^\perp)$ corresponds to the sublattice A (B).
Correspondingly, the areas for both sublattices are different from each other,
which is consistent with the existence of the sublattice imbalance.

\begin{figure*}
  \includegraphics[width=\textwidth]{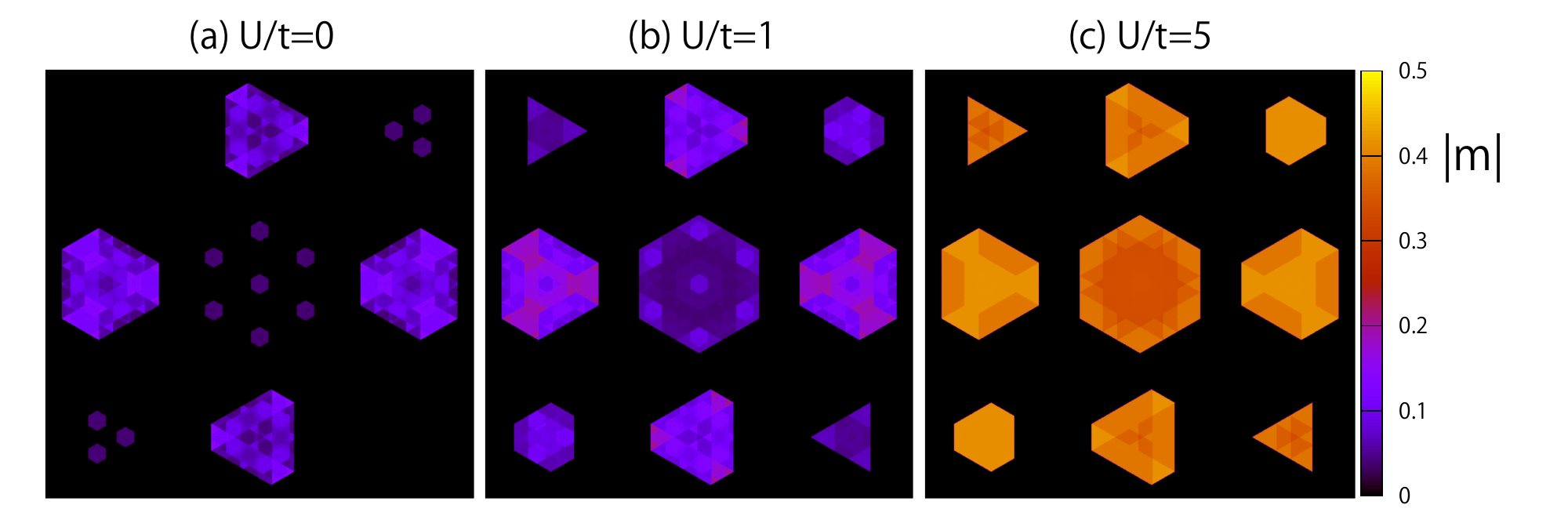}
  \caption{
    Magnetization profile in the perpendicular space $(\tilde{x},\tilde{y})$
    for the Hubbard model with $N=448\,213$
    when (a) $U/t=1.0\times 10^{-6}$, (b) $U/t=1$, and (c) $U/t=5$.
    }
  \label{perp2}
\end{figure*}

The magnetization profile in perpendicular space is shown in Fig.~\ref{perp2}, where we have shown the absolute values of the local magnetizations.
When $U/t=1.0\times 10^{-6}$,
the finite magnetization appears in the whole of
the planes $(0,\pm 1)$ and $(\pm 1,0)$,
implying that the local magnetizations appear in the sublattice B.
In contrast, in the sublattice A,
finite magnetization appears only in specific hexagonal regions
in three planes: $(-1, -1)$, $(0,0)$, and $(1, 1)$.
This is consistent with the fact that
only the confined states $\Psi_1$ with amplitudes
at C$_4$, D$_8$, and F$_0$ vertices are magnetized.
Upon increasing the interaction strength, all vertex sites are magnetized,
as shown in Fig.~\ref{perp2}(b).
When $U/t=5$, the local magnetization takes large values.
An important point is that the magnetization profile in each plane
is different from that for the weak coupling case.
In the weak coupling case, the magnetization profile originates from
the spatial distribution of the confined states.
On the other hand, in the strong coupling case,
the Coulomb interactions play a crucial role in stabilizing
the ferrimagnetically ordered states,
where intersite correlations become important.
In fact, the magnitude of local magnetizations
can be classified into four groups specified by the coordination number,
as shown in Fig.~\ref{perp2}(c) by the four separate groups of colour contrast regions.

Before concluding,
we would like to comment and compare the magnetic properties derived in the Hubbard model
on the Penrose, Ammann-Beenker, Socolar dodecagonal, and hexagonal golden-mean tilings.
One of the common features across these tilings is the existence of confined states at $E=0$
in the noninteracting case $(U=0)$,
which play a crucial role in stabilizing the magnetically ordered states
in the weak coupling limit.
Nevertheless, their confined state properties are distinct from each other.
As for the number of types of the confined states,
it is restricted to be six in the Penrose case~\cite{KohmotoSutherland,Arai},
while it should be infinite in the others.
The sublattice imbalance appears only in the hexagonal golden-mean tiling, uniquely
leading to a ferrimagnetically ordered state.
In contrast, for the other tilings, antiferromagnetically ordered states are realized
without spontaneous uniform magnetization.

\section{Summary}\label{sec:summmary}
We have studied the magnetic properties in the half-filled Hubbard model
on the hexagonal golden-mean tiling.
Examining the lattice structure, we have found that the vertex model is bipartite,
with a sublattice imbalance of $\sqrt{5}/(6\tau^3)$.
We have found the delta-function peak in the density of states of the tightbinding model,
implying the existence of macroscopically degenerate confined states at $E=0$.
We have then clarified that one type of the confined states has amplitudes in the sublattice A, while the others are only in sublattice B.
These facts give us the lower bound on the fraction of the confined states as
$(\tau+9)/(6\tau^6)\sim 0.0986$,
which is conjectured to be the exact fraction.
Furthermore, these findings lead to a ferrimagnetically ordered state even in the weak coupling limit.
Then, the introduction of the Coulomb interaction lifts the macroscopic degeneracy at the Fermi level
and induces the finite staggered magnetization as well as uniform magnetization.
By applying the real-space mean-field approach to the Hubbard model,
we have clarified how the spatial distribution of the magnetizations continuously changes
with increasing the interaction strength.
The crossover behavior in the magnetically ordered states has been discussed,
by applying the perpendicular space analysis to the local magnetizations.

\begin{acknowledgments}
  Parts of the numerical calculations are performed
  in the supercomputing systems in ISSP, the University of Tokyo.
  This work was supported by Grant-in-Aid for Scientific Research from
  JSPS, KAKENHI Grant Nos.
  JP19H05821, JP18K04678, JP17K05536 (A.K.) and JP19H05817, JP19H05818 (S.C).
\end{acknowledgments}

\bibliography{./refs}

\begin{thebibliography}{40}%
\makeatletter
\providecommand \@ifxundefined [1]{%
 \@ifx{#1\undefined}
}%
\providecommand \@ifnum [1]{%
 \ifnum #1\expandafter \@firstoftwo
 \else \expandafter \@secondoftwo
 \fi
}%
\providecommand \@ifx [1]{%
 \ifx #1\expandafter \@firstoftwo
 \else \expandafter \@secondoftwo
 \fi
}%
\providecommand \natexlab [1]{#1}%
\providecommand \enquote  [1]{``#1''}%
\providecommand \bibnamefont  [1]{#1}%
\providecommand \bibfnamefont [1]{#1}%
\providecommand \citenamefont [1]{#1}%
\providecommand \href@noop [0]{\@secondoftwo}%
\providecommand \href [0]{\begingroup \@sanitize@url \@href}%
\providecommand \@href[1]{\@@startlink{#1}\@@href}%
\providecommand \@@href[1]{\endgroup#1\@@endlink}%
\providecommand \@sanitize@url [0]{\catcode `\\12\catcode `\$12\catcode
  `\&12\catcode `\#12\catcode `\^12\catcode `\_12\catcode `\%12\relax}%
\providecommand \@@startlink[1]{}%
\providecommand \@@endlink[0]{}%
\providecommand \url  [0]{\begingroup\@sanitize@url \@url }%
\providecommand \@url [1]{\endgroup\@href {#1}{\urlprefix }}%
\providecommand \urlprefix  [0]{URL }%
\providecommand \Eprint [0]{\href }%
\providecommand \doibase [0]{http://dx.doi.org/}%
\providecommand \selectlanguage [0]{\@gobble}%
\providecommand \bibinfo  [0]{\@secondoftwo}%
\providecommand \bibfield  [0]{\@secondoftwo}%
\providecommand \translation [1]{[#1]}%
\providecommand \BibitemOpen [0]{}%
\providecommand \bibitemStop [0]{}%
\providecommand \bibitemNoStop [0]{.\EOS\space}%
\providecommand \EOS [0]{\spacefactor3000\relax}%
\providecommand \BibitemShut  [1]{\csname bibitem#1\endcsname}%
\let\auto@bib@innerbib\@empty
\bibitem [{\citenamefont {Shechtman}\ \emph {et~al.}(1984)\citenamefont
  {Shechtman}, \citenamefont {Blech}, \citenamefont {Gratias},\ and\
  \citenamefont {Cahn}}]{Shechtman}%
  \BibitemOpen
  \bibfield  {author} {\bibinfo {author} {\bibfnamefont {D.}~\bibnamefont
  {Shechtman}}, \bibinfo {author} {\bibfnamefont {I.}~\bibnamefont {Blech}},
  \bibinfo {author} {\bibfnamefont {D.}~\bibnamefont {Gratias}}, \ and\
  \bibinfo {author} {\bibfnamefont {J.~W.}\ \bibnamefont {Cahn}},\ }\href
  {\doibase 10.1103/PhysRevLett.53.1951} {\bibfield  {journal} {\bibinfo
  {journal} {Phys. Rev. Lett.}\ }\textbf {\bibinfo {volume} {53}},\ \bibinfo
  {pages} {1951} (\bibinfo {year} {1984})}\BibitemShut {NoStop}%
\bibitem [{\citenamefont {Ishimasa}\ \emph {et~al.}(2011)\citenamefont
  {Ishimasa}, \citenamefont {Tanaka},\ and\ \citenamefont
  {Kashimoto}}]{Ishimasa_2011}%
  \BibitemOpen
  \bibfield  {author} {\bibinfo {author} {\bibfnamefont {T.}~\bibnamefont
  {Ishimasa}}, \bibinfo {author} {\bibfnamefont {Y.}~\bibnamefont {Tanaka}}, \
  and\ \bibinfo {author} {\bibfnamefont {S.}~\bibnamefont {Kashimoto}},\ }\href
  {\doibase 10.1080/14786435.2011.608732} {\bibfield  {journal} {\bibinfo
  {journal} {Phil. Mag.}\ }\textbf {\bibinfo {volume} {91}},\ \bibinfo {pages}
  {4218} (\bibinfo {year} {2011})}\BibitemShut {NoStop}%
\bibitem [{\citenamefont {Deguchi}\ \emph {et~al.}(2012)\citenamefont
  {Deguchi}, \citenamefont {Matsukawa}, \citenamefont {Sato}, \citenamefont
  {Hattori}, \citenamefont {Ishida}, \citenamefont {Takakura},\ and\
  \citenamefont {Ishimasa}}]{Deguchi_2012}%
  \BibitemOpen
  \bibfield  {author} {\bibinfo {author} {\bibfnamefont {K.}~\bibnamefont
  {Deguchi}}, \bibinfo {author} {\bibfnamefont {S.}~\bibnamefont {Matsukawa}},
  \bibinfo {author} {\bibfnamefont {N.~K.}\ \bibnamefont {Sato}}, \bibinfo
  {author} {\bibfnamefont {T.}~\bibnamefont {Hattori}}, \bibinfo {author}
  {\bibfnamefont {K.}~\bibnamefont {Ishida}}, \bibinfo {author} {\bibfnamefont
  {H.}~\bibnamefont {Takakura}}, \ and\ \bibinfo {author} {\bibfnamefont
  {T.}~\bibnamefont {Ishimasa}},\ }\href {\doibase 10.1038/nmat3432} {\bibfield
   {journal} {\bibinfo  {journal} {Nat. Mat.}\ }\textbf {\bibinfo {volume}
  {11}},\ \bibinfo {pages} {1013} (\bibinfo {year} {2012})}\BibitemShut
  {NoStop}%
\bibitem [{\citenamefont {Kamiya}\ \emph {et~al.}(2018)\citenamefont {Kamiya},
  \citenamefont {Takeuchi}, \citenamefont {Kabeya}, \citenamefont {Wada},
  \citenamefont {Ishimasa}, \citenamefont {Ochiai}, \citenamefont {Deguchi},
  \citenamefont {Imura},\ and\ \citenamefont {Sato}}]{Kamiya_2018}%
  \BibitemOpen
  \bibfield  {author} {\bibinfo {author} {\bibfnamefont {K.}~\bibnamefont
  {Kamiya}}, \bibinfo {author} {\bibfnamefont {T.}~\bibnamefont {Takeuchi}},
  \bibinfo {author} {\bibfnamefont {N.}~\bibnamefont {Kabeya}}, \bibinfo
  {author} {\bibfnamefont {N.}~\bibnamefont {Wada}}, \bibinfo {author}
  {\bibfnamefont {T.}~\bibnamefont {Ishimasa}}, \bibinfo {author}
  {\bibfnamefont {A.}~\bibnamefont {Ochiai}}, \bibinfo {author} {\bibfnamefont
  {K.}~\bibnamefont {Deguchi}}, \bibinfo {author} {\bibfnamefont
  {K.}~\bibnamefont {Imura}}, \ and\ \bibinfo {author} {\bibfnamefont {N.~K.}\
  \bibnamefont {Sato}},\ }\href {\doibase 10.1038/s41467-017-02667-x}
  {\bibfield  {journal} {\bibinfo  {journal} {Nat. Comm.}\ }\textbf {\bibinfo
  {volume} {9}},\ \bibinfo {pages} {154} (\bibinfo {year} {2018})}\BibitemShut
  {NoStop}%
\bibitem [{\citenamefont {Kimura}\ \emph {et~al.}(1986)\citenamefont {Kimura},
  \citenamefont {Hashimoto}, \citenamefont {Suzuki}, \citenamefont {Nagayama},
  \citenamefont {Ino},\ and\ \citenamefont {Takeuchi}}]{Kimura_1986}%
  \BibitemOpen
  \bibfield  {author} {\bibinfo {author} {\bibfnamefont {K.}~\bibnamefont
  {Kimura}}, \bibinfo {author} {\bibfnamefont {T.}~\bibnamefont {Hashimoto}},
  \bibinfo {author} {\bibfnamefont {K.}~\bibnamefont {Suzuki}}, \bibinfo
  {author} {\bibfnamefont {K.}~\bibnamefont {Nagayama}}, \bibinfo {author}
  {\bibfnamefont {H.}~\bibnamefont {Ino}}, \ and\ \bibinfo {author}
  {\bibfnamefont {S.}~\bibnamefont {Takeuchi}},\ }\href {\doibase
  10.1143/JPSJ.55.534} {\bibfield  {journal} {\bibinfo  {journal} {J. Phys.
  Soc. Jpn.}\ }\textbf {\bibinfo {volume} {55}},\ \bibinfo {pages} {534}
  (\bibinfo {year} {1986})}\BibitemShut {NoStop}%
\bibitem [{\citenamefont {Hauser}\ \emph {et~al.}(1986)\citenamefont {Hauser},
  \citenamefont {Chen},\ and\ \citenamefont {Waszczak}}]{SG1}%
  \BibitemOpen
  \bibfield  {author} {\bibinfo {author} {\bibfnamefont {J.~J.}\ \bibnamefont
  {Hauser}}, \bibinfo {author} {\bibfnamefont {H.~S.}\ \bibnamefont {Chen}}, \
  and\ \bibinfo {author} {\bibfnamefont {J.~V.}\ \bibnamefont {Waszczak}},\
  }\href {\doibase 10.1103/PhysRevB.33.3577} {\bibfield  {journal} {\bibinfo
  {journal} {Phys. Rev. B}\ }\textbf {\bibinfo {volume} {33}},\ \bibinfo
  {pages} {3577} (\bibinfo {year} {1986})}\BibitemShut {NoStop}%
\bibitem [{\citenamefont {Fukamichi}\ \emph {et~al.}(1987)\citenamefont
  {Fukamichi}, \citenamefont {Goto}, \citenamefont {Masumoto}, \citenamefont
  {Sakakibara}, \citenamefont {Oguchi},\ and\ \citenamefont {Todo}}]{SG2}%
  \BibitemOpen
  \bibfield  {author} {\bibinfo {author} {\bibfnamefont {K.}~\bibnamefont
  {Fukamichi}}, \bibinfo {author} {\bibfnamefont {T.}~\bibnamefont {Goto}},
  \bibinfo {author} {\bibfnamefont {T.}~\bibnamefont {Masumoto}}, \bibinfo
  {author} {\bibfnamefont {T.}~\bibnamefont {Sakakibara}}, \bibinfo {author}
  {\bibfnamefont {M.}~\bibnamefont {Oguchi}}, \ and\ \bibinfo {author}
  {\bibfnamefont {S.}~\bibnamefont {Todo}},\ }\href {\doibase
  10.1088/0305-4608/17/3/018} {\bibfield  {journal} {\bibinfo  {journal} {J.
  Physics F: Met. Phys.}\ }\textbf {\bibinfo {volume} {17}},\ \bibinfo {pages}
  {743} (\bibinfo {year} {1987})}\BibitemShut {NoStop}%
\bibitem [{\citenamefont {Chen}\ and\ \citenamefont {Chen}(1986)}]{Chen_1986}%
  \BibitemOpen
  \bibfield  {author} {\bibinfo {author} {\bibfnamefont {C.~H.}\ \bibnamefont
  {Chen}}\ and\ \bibinfo {author} {\bibfnamefont {H.~S.}\ \bibnamefont
  {Chen}},\ }\href {\doibase 10.1103/PhysRevB.33.2814} {\bibfield  {journal}
  {\bibinfo  {journal} {Phys. Rev. B}\ }\textbf {\bibinfo {volume} {33}},\
  \bibinfo {pages} {2814} (\bibinfo {year} {1986})}\BibitemShut {NoStop}%
\bibitem [{\citenamefont {Tsai}\ \emph {et~al.}(1988)\citenamefont {Tsai},
  \citenamefont {Inoue}, \citenamefont {Masumoto},\ and\ \citenamefont
  {Kataoka}}]{Tsai_1988}%
  \BibitemOpen
  \bibfield  {author} {\bibinfo {author} {\bibfnamefont {A.-P.}\ \bibnamefont
  {Tsai}}, \bibinfo {author} {\bibfnamefont {A.}~\bibnamefont {Inoue}},
  \bibinfo {author} {\bibfnamefont {T.}~\bibnamefont {Masumoto}}, \ and\
  \bibinfo {author} {\bibfnamefont {N.}~\bibnamefont {Kataoka}},\ }\href
  {\doibase 10.1143/JJAP.27.L2252} {\bibfield  {journal} {\bibinfo  {journal}
  {Jpn. J. Appl. Phys.}\ }\textbf {\bibinfo {volume} {27}},\ \bibinfo {pages}
  {L2252} (\bibinfo {year} {1988})}\BibitemShut {NoStop}%
\bibitem [{\citenamefont {Hundley}\ \emph {et~al.}(1992)\citenamefont
  {Hundley}, \citenamefont {Mchenry}, \citenamefont {Dunlap}, \citenamefont
  {Srinivas},\ and\ \citenamefont {Bahadur}}]{AlMnGe}%
  \BibitemOpen
  \bibfield  {author} {\bibinfo {author} {\bibfnamefont {M.~F.}\ \bibnamefont
  {Hundley}}, \bibinfo {author} {\bibfnamefont {M.~E.}\ \bibnamefont
  {Mchenry}}, \bibinfo {author} {\bibfnamefont {R.~A.}\ \bibnamefont {Dunlap}},
  \bibinfo {author} {\bibfnamefont {V.}~\bibnamefont {Srinivas}}, \ and\
  \bibinfo {author} {\bibfnamefont {D.}~\bibnamefont {Bahadur}},\ }\href
  {\doibase 10.1080/13642819208224587} {\bibfield  {journal} {\bibinfo
  {journal} {Phil.o Mag. B}\ }\textbf {\bibinfo {volume} {66}},\ \bibinfo
  {pages} {239} (\bibinfo {year} {1992})}\BibitemShut {NoStop}%
\bibitem [{\citenamefont {Hattori}\ \emph {et~al.}(1995)\citenamefont
  {Hattori}, \citenamefont {Niikura}, \citenamefont {Tsai}, \citenamefont
  {Inoue}, \citenamefont {Masumoto}, \citenamefont {Fukamichi}, \citenamefont
  {Aruga-Katori},\ and\ \citenamefont {Goto}}]{Hattori_1995}%
  \BibitemOpen
  \bibfield  {author} {\bibinfo {author} {\bibfnamefont {Y.}~\bibnamefont
  {Hattori}}, \bibinfo {author} {\bibfnamefont {A.}~\bibnamefont {Niikura}},
  \bibinfo {author} {\bibfnamefont {A.~P.}\ \bibnamefont {Tsai}}, \bibinfo
  {author} {\bibfnamefont {A.}~\bibnamefont {Inoue}}, \bibinfo {author}
  {\bibfnamefont {T.}~\bibnamefont {Masumoto}}, \bibinfo {author}
  {\bibfnamefont {K.}~\bibnamefont {Fukamichi}}, \bibinfo {author}
  {\bibfnamefont {H.}~\bibnamefont {Aruga-Katori}}, \ and\ \bibinfo {author}
  {\bibfnamefont {T.}~\bibnamefont {Goto}},\ }\href {\doibase
  10.1088/0953-8984/7/11/009} {\bibfield  {journal} {\bibinfo  {journal} {J.
  Phys.: Condens. Matter}\ }\textbf {\bibinfo {volume} {7}},\ \bibinfo {pages}
  {2313} (\bibinfo {year} {1995})}\BibitemShut {NoStop}%
\bibitem [{\citenamefont {Charrier}\ and\ \citenamefont
  {Schmitt}(1997)}]{Charrier_1997}%
  \BibitemOpen
  \bibfield  {author} {\bibinfo {author} {\bibfnamefont {B.}~\bibnamefont
  {Charrier}}\ and\ \bibinfo {author} {\bibfnamefont {D.}~\bibnamefont
  {Schmitt}},\ }\href {\doibase 10.1016/S0304-8853(97)00062-0} {\bibfield
  {journal} {\bibinfo  {journal} {J. Mag. Mag. Mat.}\ }\textbf {\bibinfo
  {volume} {171}},\ \bibinfo {pages} {106} (\bibinfo {year}
  {1997})}\BibitemShut {NoStop}%
\bibitem [{\citenamefont {Sato}\ \emph {et~al.}(1998)\citenamefont {Sato},
  \citenamefont {Takakura}, \citenamefont {Tsai},\ and\ \citenamefont
  {Shibata}}]{Sato_1998}%
  \BibitemOpen
  \bibfield  {author} {\bibinfo {author} {\bibfnamefont {T.~J.}\ \bibnamefont
  {Sato}}, \bibinfo {author} {\bibfnamefont {H.}~\bibnamefont {Takakura}},
  \bibinfo {author} {\bibfnamefont {A.~P.}\ \bibnamefont {Tsai}}, \ and\
  \bibinfo {author} {\bibfnamefont {K.}~\bibnamefont {Shibata}},\ }\href
  {\doibase 10.1103/PhysRevLett.81.2364} {\bibfield  {journal} {\bibinfo
  {journal} {Phys. Rev. Lett.}\ }\textbf {\bibinfo {volume} {81}},\ \bibinfo
  {pages} {2364} (\bibinfo {year} {1998})}\BibitemShut {NoStop}%
\bibitem [{\citenamefont {Sato}\ \emph {et~al.}(2001)\citenamefont {Sato},
  \citenamefont {Guo},\ and\ \citenamefont {Tsai}}]{Sato_2001}%
  \BibitemOpen
  \bibfield  {author} {\bibinfo {author} {\bibfnamefont {T.~J.}\ \bibnamefont
  {Sato}}, \bibinfo {author} {\bibfnamefont {J.}~\bibnamefont {Guo}}, \ and\
  \bibinfo {author} {\bibfnamefont {A.~P.}\ \bibnamefont {Tsai}},\ }\href
  {\doibase 10.1088/0953-8984/13/4/106} {\bibfield  {journal} {\bibinfo
  {journal} {J. Phys.: Condensed Matter}\ }\textbf {\bibinfo {volume} {13}},\
  \bibinfo {pages} {L105} (\bibinfo {year} {2001})}\BibitemShut {NoStop}%
\bibitem [{\citenamefont {Tamura}\ \emph {et~al.}(2010)\citenamefont {Tamura},
  \citenamefont {Muro}, \citenamefont {Hiroto}, \citenamefont {Nishimoto},\
  and\ \citenamefont {Takabatake}}]{Tamura_2010}%
  \BibitemOpen
  \bibfield  {author} {\bibinfo {author} {\bibfnamefont {R.}~\bibnamefont
  {Tamura}}, \bibinfo {author} {\bibfnamefont {Y.}~\bibnamefont {Muro}},
  \bibinfo {author} {\bibfnamefont {T.}~\bibnamefont {Hiroto}}, \bibinfo
  {author} {\bibfnamefont {K.}~\bibnamefont {Nishimoto}}, \ and\ \bibinfo
  {author} {\bibfnamefont {T.}~\bibnamefont {Takabatake}},\ }\href {\doibase
  10.1103/PhysRevB.82.220201} {\bibfield  {journal} {\bibinfo  {journal} {Phys.
  Rev. B}\ }\textbf {\bibinfo {volume} {82}},\ \bibinfo {pages} {220201(R)}
  (\bibinfo {year} {2010})}\BibitemShut {NoStop}%
\bibitem [{\citenamefont {Tamura}\ \emph {et~al.}(2021)\citenamefont {Tamura},
  \citenamefont {Ishikawa}, \citenamefont {Suzuki}, \citenamefont {Kotajima},
  \citenamefont {Tanaka}, \citenamefont {Seki}, \citenamefont {Shibata},
  \citenamefont {Yamada}, \citenamefont {Fujii}, \citenamefont {Wang},
  \citenamefont {Avdeev}, \citenamefont {Nawa}, \citenamefont {Okuyama},\ and\
  \citenamefont {Sato}}]{Tamura}%
  \BibitemOpen
  \bibfield  {author} {\bibinfo {author} {\bibfnamefont {R.}~\bibnamefont
  {Tamura}}, \bibinfo {author} {\bibfnamefont {A.}~\bibnamefont {Ishikawa}},
  \bibinfo {author} {\bibfnamefont {S.}~\bibnamefont {Suzuki}}, \bibinfo
  {author} {\bibfnamefont {T.}~\bibnamefont {Kotajima}}, \bibinfo {author}
  {\bibfnamefont {Y.}~\bibnamefont {Tanaka}}, \bibinfo {author} {\bibfnamefont
  {T.}~\bibnamefont {Seki}}, \bibinfo {author} {\bibfnamefont {N.}~\bibnamefont
  {Shibata}}, \bibinfo {author} {\bibfnamefont {T.}~\bibnamefont {Yamada}},
  \bibinfo {author} {\bibfnamefont {T.}~\bibnamefont {Fujii}}, \bibinfo
  {author} {\bibfnamefont {C.-W.}\ \bibnamefont {Wang}}, \bibinfo {author}
  {\bibfnamefont {M.}~\bibnamefont {Avdeev}}, \bibinfo {author} {\bibfnamefont
  {K.}~\bibnamefont {Nawa}}, \bibinfo {author} {\bibfnamefont {D.}~\bibnamefont
  {Okuyama}}, \ and\ \bibinfo {author} {\bibfnamefont {T.~J.}\ \bibnamefont
  {Sato}},\ }\href {\doibase 10.1021/jacs.1c09954} {\bibfield  {journal}
  {\bibinfo  {journal} {J. Ame. Chem. Soc.}\ }\textbf {\bibinfo {volume}
  {143}},\ \bibinfo {pages} {19938} (\bibinfo {year} {2021})}\BibitemShut
  {NoStop}%
\bibitem [{\citenamefont {Takemori}\ and\ \citenamefont
  {Koga}(2015)}]{Takemori}%
  \BibitemOpen
  \bibfield  {author} {\bibinfo {author} {\bibfnamefont {N.}~\bibnamefont
  {Takemori}}\ and\ \bibinfo {author} {\bibfnamefont {A.}~\bibnamefont
  {Koga}},\ }\href {\doibase 10.7566/JPSJ.84.023701} {\bibfield  {journal}
  {\bibinfo  {journal} {J. Phys. Soc. Jpn.}\ }\textbf {\bibinfo {volume}
  {84}},\ \bibinfo {pages} {023701} (\bibinfo {year} {2015})}\BibitemShut
  {NoStop}%
\bibitem [{\citenamefont {Takemura}\ \emph {et~al.}(2015)\citenamefont
  {Takemura}, \citenamefont {Takemori},\ and\ \citenamefont {Koga}}]{Takemura}%
  \BibitemOpen
  \bibfield  {author} {\bibinfo {author} {\bibfnamefont {S.}~\bibnamefont
  {Takemura}}, \bibinfo {author} {\bibfnamefont {N.}~\bibnamefont {Takemori}},
  \ and\ \bibinfo {author} {\bibfnamefont {A.}~\bibnamefont {Koga}},\ }\href
  {\doibase 10.1103/PhysRevB.91.165114} {\bibfield  {journal} {\bibinfo
  {journal} {Phys. Rev. B}\ }\textbf {\bibinfo {volume} {91}},\ \bibinfo
  {pages} {165114} (\bibinfo {year} {2015})}\BibitemShut {NoStop}%
\bibitem [{\citenamefont {Shinzaki}\ \emph {et~al.}(2016)\citenamefont
  {Shinzaki}, \citenamefont {Nasu},\ and\ \citenamefont {Koga}}]{Shinzaki}%
  \BibitemOpen
  \bibfield  {author} {\bibinfo {author} {\bibfnamefont {R.}~\bibnamefont
  {Shinzaki}}, \bibinfo {author} {\bibfnamefont {J.}~\bibnamefont {Nasu}}, \
  and\ \bibinfo {author} {\bibfnamefont {A.}~\bibnamefont {Koga}},\ }\href
  {\doibase 10.7566/JPSJ.85.114706} {\bibfield  {journal} {\bibinfo  {journal}
  {J. Phys. Soc. Jpn.}\ }\textbf {\bibinfo {volume} {85}},\ \bibinfo {pages}
  {114706} (\bibinfo {year} {2016})}\BibitemShut {NoStop}%
\bibitem [{\citenamefont {Hauck}\ \emph {et~al.}(2021)\citenamefont {Hauck},
  \citenamefont {Honerkamp}, \citenamefont {Achilles},\ and\ \citenamefont
  {Kennes}}]{Hauck}%
  \BibitemOpen
  \bibfield  {author} {\bibinfo {author} {\bibfnamefont {J.~B.}\ \bibnamefont
  {Hauck}}, \bibinfo {author} {\bibfnamefont {C.}~\bibnamefont {Honerkamp}},
  \bibinfo {author} {\bibfnamefont {S.}~\bibnamefont {Achilles}}, \ and\
  \bibinfo {author} {\bibfnamefont {D.~M.}\ \bibnamefont {Kennes}},\ }\href
  {\doibase 10.1103/PhysRevResearch.3.023180} {\bibfield  {journal} {\bibinfo
  {journal} {Phys. Rev. Research}\ }\textbf {\bibinfo {volume} {3}},\ \bibinfo
  {pages} {023180} (\bibinfo {year} {2021})}\BibitemShut {NoStop}%
\bibitem [{\citenamefont {Sakai}\ \emph {et~al.}(2017)\citenamefont {Sakai},
  \citenamefont {Takemori}, \citenamefont {Koga},\ and\ \citenamefont
  {Arita}}]{Sakai_2017}%
  \BibitemOpen
  \bibfield  {author} {\bibinfo {author} {\bibfnamefont {S.}~\bibnamefont
  {Sakai}}, \bibinfo {author} {\bibfnamefont {N.}~\bibnamefont {Takemori}},
  \bibinfo {author} {\bibfnamefont {A.}~\bibnamefont {Koga}}, \ and\ \bibinfo
  {author} {\bibfnamefont {R.}~\bibnamefont {Arita}},\ }\href {\doibase
  10.1103/PhysRevB.95.024509} {\bibfield  {journal} {\bibinfo  {journal} {Phys.
  Rev. B}\ }\textbf {\bibinfo {volume} {95}},\ \bibinfo {pages} {024509}
  (\bibinfo {year} {2017})}\BibitemShut {NoStop}%
\bibitem [{\citenamefont {Sakai}\ and\ \citenamefont
  {Arita}(2019)}]{Sakai_2019}%
  \BibitemOpen
  \bibfield  {author} {\bibinfo {author} {\bibfnamefont {S.}~\bibnamefont
  {Sakai}}\ and\ \bibinfo {author} {\bibfnamefont {R.}~\bibnamefont {Arita}},\
  }\href {\doibase 10.1103/PhysRevResearch.1.022002} {\bibfield  {journal}
  {\bibinfo  {journal} {Phys. Rev. Research}\ }\textbf {\bibinfo {volume}
  {1}},\ \bibinfo {pages} {022002(R)} (\bibinfo {year} {2019})}\BibitemShut
  {NoStop}%
\bibitem [{\citenamefont {Inayoshi}\ \emph {et~al.}(2020)\citenamefont
  {Inayoshi}, \citenamefont {Murakami},\ and\ \citenamefont
  {Koga}}]{Inayoshi_2020}%
  \BibitemOpen
  \bibfield  {author} {\bibinfo {author} {\bibfnamefont {K.}~\bibnamefont
  {Inayoshi}}, \bibinfo {author} {\bibfnamefont {Y.}~\bibnamefont {Murakami}},
  \ and\ \bibinfo {author} {\bibfnamefont {A.}~\bibnamefont {Koga}},\ }\href
  {\doibase 10.7566/JPSJ.89.064002} {\bibfield  {journal} {\bibinfo  {journal}
  {J. Phys. Soc. Jpn.}\ }\textbf {\bibinfo {volume} {89}},\ \bibinfo {pages}
  {064002} (\bibinfo {year} {2020})}\BibitemShut {NoStop}%
\bibitem [{\citenamefont {Cao}\ \emph {et~al.}(2020)\citenamefont {Cao},
  \citenamefont {Zhang}, \citenamefont {Liu}, \citenamefont {Liu},
  \citenamefont {Chen},\ and\ \citenamefont {Yang}}]{Cao}%
  \BibitemOpen
  \bibfield  {author} {\bibinfo {author} {\bibfnamefont {Y.}~\bibnamefont
  {Cao}}, \bibinfo {author} {\bibfnamefont {Y.}~\bibnamefont {Zhang}}, \bibinfo
  {author} {\bibfnamefont {Y.-B.}\ \bibnamefont {Liu}}, \bibinfo {author}
  {\bibfnamefont {C.-C.}\ \bibnamefont {Liu}}, \bibinfo {author} {\bibfnamefont
  {W.-Q.}\ \bibnamefont {Chen}}, \ and\ \bibinfo {author} {\bibfnamefont
  {F.}~\bibnamefont {Yang}},\ }\href {\doibase 10.1103/PhysRevLett.125.017002}
  {\bibfield  {journal} {\bibinfo  {journal} {Phys. Rev. Lett.}\ }\textbf
  {\bibinfo {volume} {125}},\ \bibinfo {pages} {017002} (\bibinfo {year}
  {2020})}\BibitemShut {NoStop}%
\bibitem [{\citenamefont {Takemori}\ \emph {et~al.}(2020)\citenamefont
  {Takemori}, \citenamefont {Arita},\ and\ \citenamefont
  {Sakai}}]{Takemori_2020}%
  \BibitemOpen
  \bibfield  {author} {\bibinfo {author} {\bibfnamefont {N.}~\bibnamefont
  {Takemori}}, \bibinfo {author} {\bibfnamefont {R.}~\bibnamefont {Arita}}, \
  and\ \bibinfo {author} {\bibfnamefont {S.}~\bibnamefont {Sakai}},\ }\href
  {\doibase 10.1103/PhysRevB.102.115108} {\bibfield  {journal} {\bibinfo
  {journal} {Phys. Rev. B}\ }\textbf {\bibinfo {volume} {102}},\ \bibinfo
  {pages} {115108} (\bibinfo {year} {2020})}\BibitemShut {NoStop}%
\bibitem [{\citenamefont {Wessel}\ \emph {et~al.}(2003)\citenamefont {Wessel},
  \citenamefont {Jagannathan},\ and\ \citenamefont {Haas}}]{Wessel_2003}%
  \BibitemOpen
  \bibfield  {author} {\bibinfo {author} {\bibfnamefont {S.}~\bibnamefont
  {Wessel}}, \bibinfo {author} {\bibfnamefont {A.}~\bibnamefont {Jagannathan}},
  \ and\ \bibinfo {author} {\bibfnamefont {S.}~\bibnamefont {Haas}},\ }\href
  {\doibase 10.1103/PhysRevLett.90.177205} {\bibfield  {journal} {\bibinfo
  {journal} {Phys. Rev. Lett.}\ }\textbf {\bibinfo {volume} {90}},\ \bibinfo
  {pages} {177205} (\bibinfo {year} {2003})}\BibitemShut {NoStop}%
\bibitem [{\citenamefont {Jagannathan}\ \emph {et~al.}(2007)\citenamefont
  {Jagannathan}, \citenamefont {Szallas}, \citenamefont {Wessel},\ and\
  \citenamefont {Duneau}}]{Jagannathan_2007}%
  \BibitemOpen
  \bibfield  {author} {\bibinfo {author} {\bibfnamefont {A.}~\bibnamefont
  {Jagannathan}}, \bibinfo {author} {\bibfnamefont {A.}~\bibnamefont
  {Szallas}}, \bibinfo {author} {\bibfnamefont {S.}~\bibnamefont {Wessel}}, \
  and\ \bibinfo {author} {\bibfnamefont {M.}~\bibnamefont {Duneau}},\ }\href
  {\doibase 10.1103/PhysRevB.75.212407} {\bibfield  {journal} {\bibinfo
  {journal} {Phys. Rev. B}\ }\textbf {\bibinfo {volume} {75}},\ \bibinfo
  {pages} {212407} (\bibinfo {year} {2007})}\BibitemShut {NoStop}%
\bibitem [{\citenamefont {Jagannathan}\ and\ \citenamefont
  {Schulz}(1997)}]{Jagannathan_Schulz_1997}%
  \BibitemOpen
  \bibfield  {author} {\bibinfo {author} {\bibfnamefont {A.}~\bibnamefont
  {Jagannathan}}\ and\ \bibinfo {author} {\bibfnamefont {H.~J.}\ \bibnamefont
  {Schulz}},\ }\href {\doibase 10.1103/PhysRevB.55.8045} {\bibfield  {journal}
  {\bibinfo  {journal} {Phys. Rev. B}\ }\textbf {\bibinfo {volume} {55}},\
  \bibinfo {pages} {8045} (\bibinfo {year} {1997})}\BibitemShut {NoStop}%
\bibitem [{\citenamefont {Koga}\ and\ \citenamefont
  {Tsunetsugu}(2017)}]{Koga_Tsunetsugu}%
  \BibitemOpen
  \bibfield  {author} {\bibinfo {author} {\bibfnamefont {A.}~\bibnamefont
  {Koga}}\ and\ \bibinfo {author} {\bibfnamefont {H.}~\bibnamefont
  {Tsunetsugu}},\ }\href {\doibase 10.1103/PhysRevB.96.214402} {\bibfield
  {journal} {\bibinfo  {journal} {Phys. Rev. B}\ }\textbf {\bibinfo {volume}
  {96}},\ \bibinfo {pages} {214402} (\bibinfo {year} {2017})}\BibitemShut
  {NoStop}%
\bibitem [{\citenamefont {Koga}(2020)}]{Koga_AB}%
  \BibitemOpen
  \bibfield  {author} {\bibinfo {author} {\bibfnamefont {A.}~\bibnamefont
  {Koga}},\ }\href {\doibase 10.1103/PhysRevB.102.115125} {\bibfield  {journal}
  {\bibinfo  {journal} {Phys. Rev. B}\ }\textbf {\bibinfo {volume} {102}},\
  \bibinfo {pages} {115125} (\bibinfo {year} {2020})}\BibitemShut {NoStop}%
\bibitem [{\citenamefont {Koga}(2021)}]{Koga_dodeca}%
  \BibitemOpen
  \bibfield  {author} {\bibinfo {author} {\bibfnamefont {A.}~\bibnamefont
  {Koga}},\ }\href {\doibase 10.2320/matertrans.MT-MB2020003} {\bibfield
  {journal} {\bibinfo  {journal} {Mater. Trans.}\ }\textbf {\bibinfo {volume}
  {62}},\ \bibinfo {pages} {360} (\bibinfo {year} {2021})}\BibitemShut
  {NoStop}%
\bibitem [{\citenamefont {Sakai}\ and\ \citenamefont {Koga}(2021)}]{SakaiKoga}%
  \BibitemOpen
  \bibfield  {author} {\bibinfo {author} {\bibfnamefont {S.}~\bibnamefont
  {Sakai}}\ and\ \bibinfo {author} {\bibfnamefont {A.}~\bibnamefont {Koga}},\
  }\href {\doibase 10.2320/matertrans.MT-MB2020001} {\bibfield  {journal}
  {\bibinfo  {journal} {Mater. Trans.}\ }\textbf {\bibinfo {volume} {62}},\
  \bibinfo {pages} {380} (\bibinfo {year} {2021})}\BibitemShut {NoStop}%
\bibitem [{\citenamefont {Coates}\ \emph {et~al.}(shed)\citenamefont {Coates},
  \citenamefont {Lifshitz}, \citenamefont {Koga}, \citenamefont {McGrath},
  \citenamefont {Sharma},\ and\ \citenamefont {Tamura}}]{Sam}%
  \BibitemOpen
  \bibfield  {author} {\bibinfo {author} {\bibfnamefont {S.}~\bibnamefont
  {Coates}}, \bibinfo {author} {\bibfnamefont {R.}~\bibnamefont {Lifshitz}},
  \bibinfo {author} {\bibfnamefont {A.}~\bibnamefont {Koga}}, \bibinfo {author}
  {\bibfnamefont {R.}~\bibnamefont {McGrath}}, \bibinfo {author} {\bibfnamefont
  {H.~R.}\ \bibnamefont {Sharma}}, \ and\ \bibinfo {author} {\bibfnamefont
  {R.}~\bibnamefont {Tamura}},\ }\href {http://arxiv.org/abs/2201.11848}
  {\bibfield  {journal} {\bibinfo  {journal} {preprint}\ ,\ \bibinfo {pages}
  {arXiv:2201.11848}} (\bibinfo {year} {unpublished})}\BibitemShut {NoStop}%
\bibitem [{\citenamefont {Lieb}(1989)}]{Lieb}%
  \BibitemOpen
  \bibfield  {author} {\bibinfo {author} {\bibfnamefont {E.~H.}\ \bibnamefont
  {Lieb}},\ }\href {\doibase 10.1103/PhysRevLett.62.1201} {\bibfield  {journal}
  {\bibinfo  {journal} {Phys. Rev. Lett.}\ }\textbf {\bibinfo {volume} {62}},\
  \bibinfo {pages} {1201} (\bibinfo {year} {1989})}\BibitemShut {NoStop}%
\bibitem [{\citenamefont {Mielke}(1992)}]{Mielke}%
  \BibitemOpen
  \bibfield  {author} {\bibinfo {author} {\bibfnamefont {A.}~\bibnamefont
  {Mielke}},\ }\href {\doibase 10.1088/0305-4470/25/16/011} {\bibfield
  {journal} {\bibinfo  {journal} {J. Phys. A: Math. Gen.}\ }\textbf {\bibinfo
  {volume} {25}},\ \bibinfo {pages} {4335–4345} (\bibinfo {year}
  {1992})}\BibitemShut {NoStop}%
\bibitem [{\citenamefont {Tasaki}(1992)}]{Tasaki}%
  \BibitemOpen
  \bibfield  {author} {\bibinfo {author} {\bibfnamefont {H.}~\bibnamefont
  {Tasaki}},\ }\href {\doibase 10.1103/PhysRevLett.69.1608} {\bibfield
  {journal} {\bibinfo  {journal} {Phys. Rev. Lett.}\ }\textbf {\bibinfo
  {volume} {69}},\ \bibinfo {pages} {1608} (\bibinfo {year}
  {1992})}\BibitemShut {NoStop}%
\bibitem [{\citenamefont {Kusakabe}\ and\ \citenamefont
  {Aoki}(1992)}]{Kusakabe}%
  \BibitemOpen
  \bibfield  {author} {\bibinfo {author} {\bibfnamefont {K.}~\bibnamefont
  {Kusakabe}}\ and\ \bibinfo {author} {\bibfnamefont {H.}~\bibnamefont
  {Aoki}},\ }\href {\doibase 10.1143/JPSJ.61.1165} {\bibfield  {journal}
  {\bibinfo  {journal} {J. Phys. Soc. Jpn.}\ }\textbf {\bibinfo {volume}
  {61}},\ \bibinfo {pages} {1165–1168} (\bibinfo {year} {1992})}\BibitemShut
  {NoStop}%
\bibitem [{\citenamefont {Noda}\ \emph {et~al.}(2009)\citenamefont {Noda},
  \citenamefont {Koga}, \citenamefont {Kawakami},\ and\ \citenamefont
  {Pruschke}}]{Noda}%
  \BibitemOpen
  \bibfield  {author} {\bibinfo {author} {\bibfnamefont {K.}~\bibnamefont
  {Noda}}, \bibinfo {author} {\bibfnamefont {A.}~\bibnamefont {Koga}}, \bibinfo
  {author} {\bibfnamefont {N.}~\bibnamefont {Kawakami}}, \ and\ \bibinfo
  {author} {\bibfnamefont {T.}~\bibnamefont {Pruschke}},\ }\href {\doibase
  10.1103/PhysRevA.80.063622} {\bibfield  {journal} {\bibinfo  {journal} {Phys.
  Rev. A}\ }\textbf {\bibinfo {volume} {80}},\ \bibinfo {pages} {063622}
  (\bibinfo {year} {2009})}\BibitemShut {NoStop}%
\bibitem [{\citenamefont {Kohmoto}\ and\ \citenamefont
  {Sutherland}(1986)}]{KohmotoSutherland}%
  \BibitemOpen
  \bibfield  {author} {\bibinfo {author} {\bibfnamefont {M.}~\bibnamefont
  {Kohmoto}}\ and\ \bibinfo {author} {\bibfnamefont {B.}~\bibnamefont
  {Sutherland}},\ }\href {\doibase 10.1103/PhysRevLett.56.2740} {\bibfield
  {journal} {\bibinfo  {journal} {Phys. Rev. Lett.}\ }\textbf {\bibinfo
  {volume} {56}},\ \bibinfo {pages} {2740} (\bibinfo {year}
  {1986})}\BibitemShut {NoStop}%
\bibitem [{\citenamefont {Arai}\ \emph {et~al.}(1988)\citenamefont {Arai},
  \citenamefont {Tokihiro}, \citenamefont {Fujiwara},\ and\ \citenamefont
  {Kohmoto}}]{Arai}%
  \BibitemOpen
  \bibfield  {author} {\bibinfo {author} {\bibfnamefont {M.}~\bibnamefont
  {Arai}}, \bibinfo {author} {\bibfnamefont {T.}~\bibnamefont {Tokihiro}},
  \bibinfo {author} {\bibfnamefont {T.}~\bibnamefont {Fujiwara}}, \ and\
  \bibinfo {author} {\bibfnamefont {M.}~\bibnamefont {Kohmoto}},\ }\href
  {\doibase 10.1103/PhysRevB.38.1621} {\bibfield  {journal} {\bibinfo
  {journal} {Phys. Rev. B}\ }\textbf {\bibinfo {volume} {38}},\ \bibinfo
  {pages} {1621} (\bibinfo {year} {1988})}\BibitemShut {NoStop}%
\end{thebibliography}%

\end{document}